\documentclass[prb,superscriptaddress,twocolumn]{revtex4-1}
\usepackage{float}
\usepackage{placeins}
\usepackage[english]{babel}
\usepackage{amsmath}
\usepackage{graphicx,epstopdf}
\usepackage[colorlinks=true,allcolors=blue]{hyperref}

\setcitestyle{super}


\begin{document}

\title{Ab initio-based Deep-Learning Prediction of Carrier Mobility in Strongly Anharmonic Materials}
\author{Juan Zhang}
\affiliation{The NOMAD Laboratory at BIFOLD Technical University of Berlin, Franklinstr. 28-29, 10587 Berlin, Germany}
\affiliation{College of Future Information Technology and State Key Laboratory of Photovoltaic Science and Technology, Fudan University, Shanghai 200433, China}

\author{Boheng Zhao}
\affiliation{Department of Physics, Tsinghua University, Beijing 100084, China}
\affiliation{State Key Laboratory of Low Dimensional Quantum Physics, Tsinghua University, Beijing 100084, China}

\author{Yang Li}
\affiliation{Department of Physics, Tsinghua University, Beijing 100084, China}
\affiliation{State Key Laboratory of Low Dimensional Quantum Physics, Tsinghua University, Beijing 100084, China}

\author{Yong Xu}
\affiliation{Department of Physics, Tsinghua University, Beijing 100084, China}
\affiliation{State Key Laboratory of Low Dimensional Quantum Physics, Tsinghua University, Beijing 100084, China}
\affiliation{Frontier Science Center for Quantum Information, Tsinghua University, Beijing 100084, China}

\author{Hao Zhang}
\affiliation{College of Future Information Technology and State Key Laboratory of Photovoltaic Science and Technology, Fudan University, Shanghai 200433, China}

\author{Kisung Kang\thanks{Corresponding author.}}
\email{kisung.kang@jnu.ac.kr}
\affiliation{The NOMAD Laboratory at BIFOLD Technical University of Berlin, Franklinstr. 28-29, 10587 Berlin, Germany}
\affiliation{School of Materials Science and Engineering, Chonnam National University, 61186 Gwangju, Republic of Korea}

\author{Matthias Scheffler}
\affiliation{The NOMAD Laboratory at BIFOLD Technical University of Berlin, Franklinstr. 28-29, 10587 Berlin, Germany}

\begin{abstract}
Predicting charge transport in strongly anharmonic materials, particularly ultralow thermal conductors, remains a major challenge for first-principles methods.
In such systems, perturbative treatments of electron-phonon interactions and the harmonic phonon picture often break down, necessitating non-perturbative approaches.
The \textit{ab initio} Kubo--Greenwood\,(aiKG) formalism provides a rigorous framework for evaluating temperature-dependent carrier transport beyond the harmonic approximation.
Nevertheless, its practical application is computationally demanding because it requires large supercells, extensive statistical sampling, and extrapolation to the zero-frequency limit.
In this work, we introduce an artificial-intelligence\,(AI)-assisted aiKG framework that incorporates the deep-learning Hamiltonian model.
By predicting the Kohn–Sham Hamiltonian with sub-meV accuracy for supercells of up to 250 atoms, the model bypasses the costly iterative self-consistent field calculations while retaining first-principles reliability within the scope of effects captured by the training data.
Using a strongly anharmonic thermal insulator, potassium iodide\,(KI) as a benchmark system, we demonstrate that the proposed approach enables efficient simulations of electronic structure and transport properties from a large supercell.
The framework reproduces temperature-dependent carrier mobilities, spectral functions, and effective masses in close agreement with the underlying density functional theory while reducing computational cost to 10\,\%.
These results suggest that the AI-assisted aiKG framework can make non-perturbative transport calculations tractable for strongly anharmonic materials, opening a scalable route towards realistic simulations and accelerated discovery of new functional materials.
\end{abstract}

\maketitle

\section{Introduction}



Thermoelectric materials offer a promising approach to converting waste heat into electrical energy, providing an attractive pathway toward sustainable energy technologies\,\cite{snyder2008,shakouri2011,he2017}.
Achieving high thermoelectric performance requires simultaneously maintaining efficient charge transport while suppressing heat conduction, a balance that requires materials with complex lattice dynamics\,\cite{li2012,zevalkink2018}.
In recent years, particular attention has been drawn to materials exhibiting extremely low lattice thermal conductivity, typically arising from strong lattice anharmonicity\,\cite{chang2018,xia2020,knoop2020,purcell}.
Although such systems are attractive candidates for thermoelectric applications, their electrical transport properties remain computationally challenging to predict from first principles.
Large-amplitude lattice fluctuations can significantly alter the electronic structure and carrier scattering processes, making a reliable description of carrier mobility a persistent challenge for theoretical approaches.

Perturbation theory has long served as the central framework for describing quasiparticle interactions in crystalline solids, including electron-phonon\,\cite{giustino,PRB224310,babadi2017,lanzara2001,wright2016,zhang2025}
couplings and so on. 
In this picture, lattice vibrations are typically described by harmonic phonons\,\cite{baroni2001} and their renormalization\,\cite{renormalization}, and charge transport is treated via perturbative scattering processes arising from the linear response of the electronic structure to nuclear displacements\,\cite{giustino,ponce2021}.
Approaches based on the Boltzmann transport equation\,\cite{peierls} or its quantum extensions, such as the Wigner transport formalism\,\cite{simoncell}, have become standard tools for modeling carrier transport in semiconductors.
The perturbative description of charge transport, which forms the basis of most first-principles transport theories, becomes unreliable in materials with strong lattice anharmonicity\,\cite{errea2014,quan2025} because it presupposes well-defined phonon quasiparticles and weak lattice fluctuations.
In strongly anharmonic materials, commonly found in thermal insulators and promising thermoelectric candidates\,\cite{chang2018,wu2020,song2023}, higher-order nuclear interactions (3rd/4th-order anharmonic terms or even higher orders) become significantly important\,\cite{xia2020,tadano2018,feng2017,wang2021,ouyang2022}.
In such regimes, phonon scattering rates can approach or even exceed the Ioffe-Regel limit\,\cite{ioffe,beltukov2013}, signaling a breakdown of the harmonic phonon picture and exposing the intrinsic limitations of perturbative descriptions of charge transport.

To overcome these limitations, non-perturbative approaches explicitly account for atomic displacements and fluctuations, thereby capturing anharmonic effects to all orders, although their practical application remains computationally demanding.
Within this framework, transport properties can be evaluated directly from thermally sampled atomic configurations, such as from molecular dynamics\,(MD).
For example, lattice thermal conductivity can be computed using the Green--Kubo formalism\,\cite{green,kubo1957,carbogno2017,knoop2023}, while electrical transport can be described using the Kubo--Greenwood method\,\cite{greenwood1958,holst2011}.
Its first-principles approach, \textit{ab initio} Kubo--Greenwood\,(aiKG) method, has been successfully applied to strongly anharmonic systems, such as SrTiO$_{3}$ and BaTiO$_{3}$\,\cite{KG}, for which perturbative approaches may become inaccurate.
These approaches naturally incorporate the effects of thermal disorder and electron-vibration coupling beyond the perturbative regime.
However, their practical application remains severely constrained by computational cost\,\cite{KG}.
Reliable transport predictions require large supercells, long MD simulations, and extensive statistical averaging, together with careful extrapolation of optical conductivity toward the zero-frequency limit\,(DC limit).
As a result, applying such non-perturbative methods to realistic material systems remains computationally demanding, and scalable strategies to alleviate these bottlenecks are still largely missing.

Recent advances in artificial intelligence\,(AI) techniques offer a promising approach to alleviate the computational bottlenecks of non-perturbative transport calculations.
In particular, deep-learning Hamiltonian models aim to learn an effective representation of the underlying quantum-mechanical Hamiltonian directly from first-principles data\,\cite{li2022deeph, 2026deeph, qian2025, hegde2017,schutt2019}.
By predicting the Kohn--Sham Hamiltonian for new atomic configurations, these models bypass the computationally demanding self-consistent field\,(SCF) iterations required in conventional density functional theory\,(DFT) calculations, while retaining first-principles accuracy.
This capability enables efficient electronic-structure evaluations for large supercells with thermally disordered configurations that would otherwise be computationally prohibitive.

In this work, we introduce an AI-assisted aiKG framework added on the \textit{ab initio} electronic-structure package \texttt{FHI-aims}\,\cite{blum2009ab,2015hybrid}.
The framework integrates the deep-learning Hamiltonian model \texttt{DeepH}\,\cite{li2022deeph,gong2023}, which is trained to predict the Kohn--Sham Hamiltonian with sub-meV accuracy for supercells of up to 250 atoms.
By replacing repeated SCF steps in electronic-structure calculations with efficient Hamiltonian predictions, this approach enables large-scale evaluations of temperature-dependent electronic structures and transport properties within the aiKG formalism.
Using potassium iodide (KI)\,\cite{dolling1966,freville2023} as a example for strongly anharmonic material, we demonstrate the predictive capability of this framework for calculating electronic band structures and carrier mobilities in large supercells.
Beyond transport coefficients, we further investigate non-perturbative temperature-dependent electronic spectral functions\,\cite{zacharias2020,quan2025} and the evaluation of hole effective masses in the presence of strong lattice fluctuations.
These results highlight the potential of AI-assisted electronic-structure methods for enabling realistic simulations of charge transport in strongly anharmonic materials. We present this work as a methodological demonstration, broader benchmarking across structurally and chemically distinct anharmonic systems is left to follow-up studies.

\section{Methodology}
\label{sec:method}

\subsection{The Deep-learning Hamiltonian added on FHI-aims}

\begin{figure}[t]
\centering
\includegraphics[width=1\linewidth]{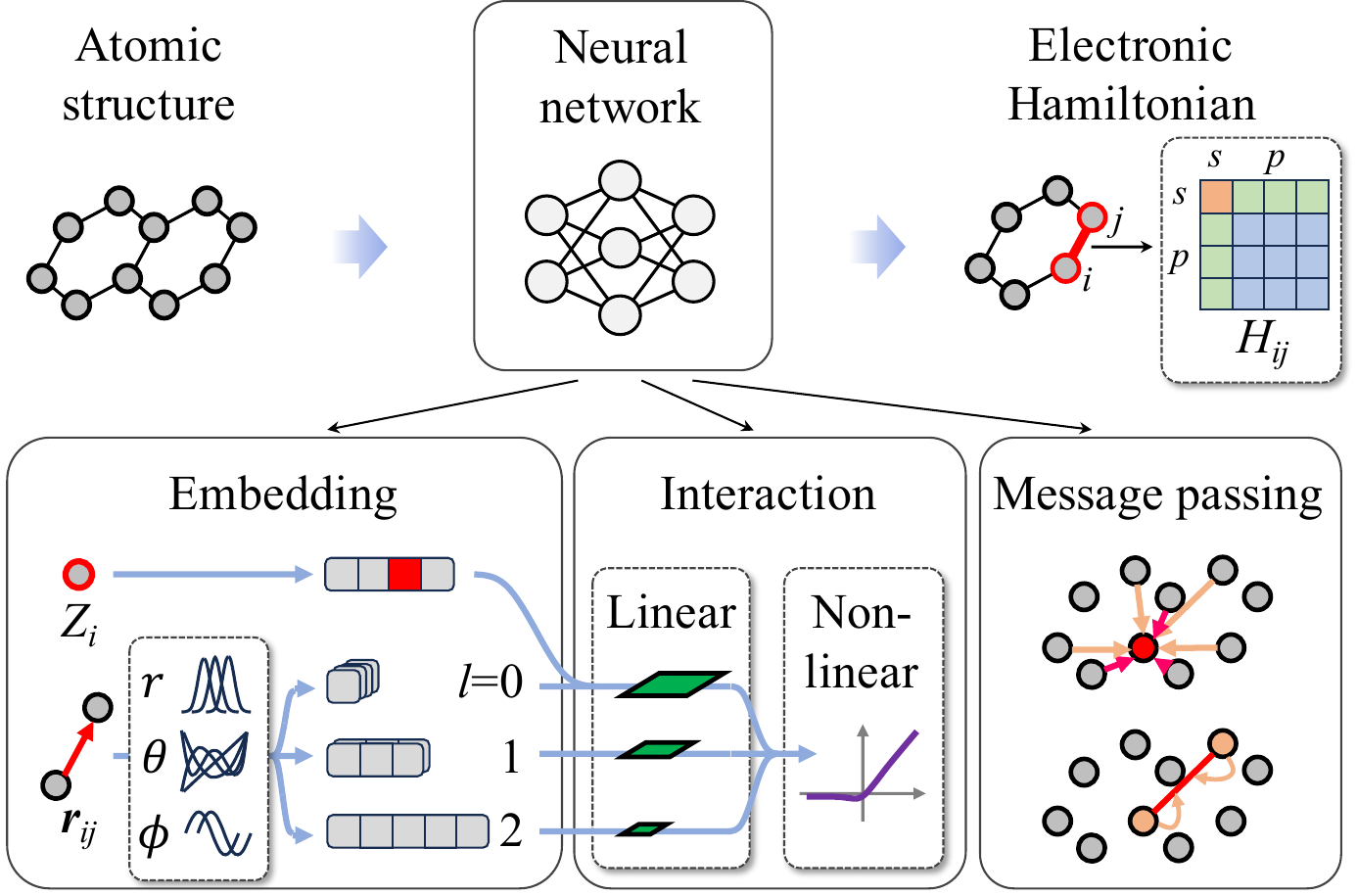}
\caption{Illustration of the framework of the DeepH method. DeepH maps the atomic structure to the electronic Hamiltonian using a graph neural network that consists of an embedding layer, several interaction layers, and several message-passing layers.}
\label{fig:deeph}
\end{figure}

Electronic-structure calculations in this work are accelerated by \texttt{DeepH}\,\cite{li2022deeph,2026deeph}, a deep-learning Hamiltonian approach that predicts the Kohn--Sham Hamiltonian directly from atomic configurations, as shown in Fig.\,~\ref{fig:deeph}. 
The effectiveness of this approach relies on the nearsightedness of electronic structure\,\cite{prodan2005}, whereby the Hamiltonian is primarily determined by the local atomic environment and depends only weakly on distant atoms.
This locality principle enables the mapping from atomic structures to the electronic Hamiltonian to be learned using graph neural networks.
In practice, the use of localized atomic orbital basis sets yields a sparse Hamiltonian matrix, thereby reducing the effective dimensionality and enabling efficient treatment of large-scale systems.
Atomic configurations are represented as geometric graphs, with nodes associated with atoms and edges connecting atom pairs that have non-zero Hamiltonian matrix elements\,\cite{li2022deeph}.
The model takes such a graph as input and performs message passing between node and edge features. After several blocks of information exchange and non-linear transformations, the resulting features exhibit strong representation capacity for describing complex many-body interactions.
Finally, the Hamiltonian matrix is constructed from the output features of the last block.
Once trained on first-principles reference data, the model can predict the electronic Hamiltonian for unseen atomic configurations without performing SCF iterations, enabling the evaluation of electronic band structures and related properties with first-principles accuracy.
The finite message-passing range enables the model to achieve linear scaling with system size, therefore significantly enhancing the calculation efficiency on large systems.
It should be noted that DeepH models can only generalize to the systems with similar local atomic configurations to the train set. To accurately predict the effects that only occur in large-scale systems, such as charge transfer and defect interactions, the train set needs carefully designed structures with relatively large supercells.

An important aspect of neural network design is to preserve the symmetries of the underlying physical principles.
The Hamiltonian matrix transforms equivariantly under translations and rotations, calling for an equivariant neural network architecture\,\cite{2026deeph}.
The \texttt{DeepH} method fully exploits physical priors to enhance the generalization ability and data efficiency of the model.
To preserve translational invariance, the model embeds the relative coordinates of atom pairs rather than the absolute coordinates of individual atoms.
The atomic structures are embedded by a proper set of bases. Namely, the atomic number $Z_i$ of each atom in the structure is embedded by one-hot vector as below:
\begin{equation}
v_{i, \mu}= \begin{cases}1, & \mu=Z_i \\ 0, & \mu \neq Z_i\end{cases}
\end{equation}
where $\mu$ scans all the atomic numbers contained in the data set. And the bond length $r_{ij}$ of each atom pair is embedded by Gaussian radial basis:
\begin{equation}
e_{i j, v}=e^{-\frac{\left(r_{i j}-r_v\right)^2}{2 \sigma^2}}
\end{equation}
where $r_v$ forms an arithmetic sequence from 0 to a cutoff radius $r_{cut}$, and $\sigma$ is the smearing width of the Gaussian functions which is set to the spacing between adjacent $r_v$. 
The direction of the bonds is taken into account through spherical harmonic embeddings labeled by the angular quantum number $l$:
\begin{equation}
e_{i j}^{l m}=Y_{l m\left(\hat{r}_{i j}\right)}
\end{equation}
and $\hat{\boldsymbol{r}}_{i j}=\frac{r_{i j}}{r_{i j}}$ denotes the unit vector along the bond direction. 
This makes the features in the neural network carry irreducible representations of the $SO(3)$ group and form a closed set under rotations\,\cite{gong2023}.
The scalar embeddings and spherical harmonic embeddings are combined to generate initial equivariant features in the neural network.
The hidden blocks in the model follow the coupling rule of angular momentum to exchange information between channels of different $l$ while preserving rotational equivariance.
To apply non-linear transformations to the equivariant features, activation functions operate on scalar features, and the activated scalars are then attached to other features with $l>0$.
After sufficient feature evolution, tensor products of equivariant feature vectors are performed via the Wigner-Eckart theorem to represent the matrix form of the electronic Hamiltonian, making the Hamiltonian matrix transform equivariantly in both the row and column dimensions.

The neural network architecture integrates a set of advanced algorithms for improved accuracy and efficiency.
Specifically, in the tensor product block that couples different angular momenta, the $SO(3)$-equivariant edge features are transformed into a semi-local coordinate frame whose $z$ axis is aligned with the bond vector $\hat{\bm{r}}_{ij}$.
The structure-adaptive rotation reduces the $SO(3)$ representations into $SO(2)$ equivariant components. This facilitates the interaction of features across different orbital angular momenta $l$ while preserving the magnetic quantum number $m$, which is algorithmically equivalent to the $SO(3)$ tensor product but is significantly more efficient\,\cite{passaro2023}.
Many other operations can be performed in the semi-local coordinate frame, such as equivariant activations and scalar attachments. Because the structure of the $SO(2)$ group is much simpler than that of the $SO(3)$ group, the operations in the semi-local frame are subject to fewer constraints and exhibit stronger expressive power.
Furthermore, the model employs an attention mechanism to modulate the message passing from edges to nodes.
The attention weights are generated by a softmax-normalized feedforward net taking edge features as input, allowing the neural network to capture the most important interactions between the central atom and the surrounding atoms, which dominate the relationship between the atomic structures and the electronic Hamiltonian.
In summary, these architectural designs enable the model to achieve strong representation capacity within physical constraints, ensuring accurate and physics-informed modeling of electronic structures across a wide range of materials.

\begin{figure}[t]
\centering
\includegraphics[width=1\linewidth]{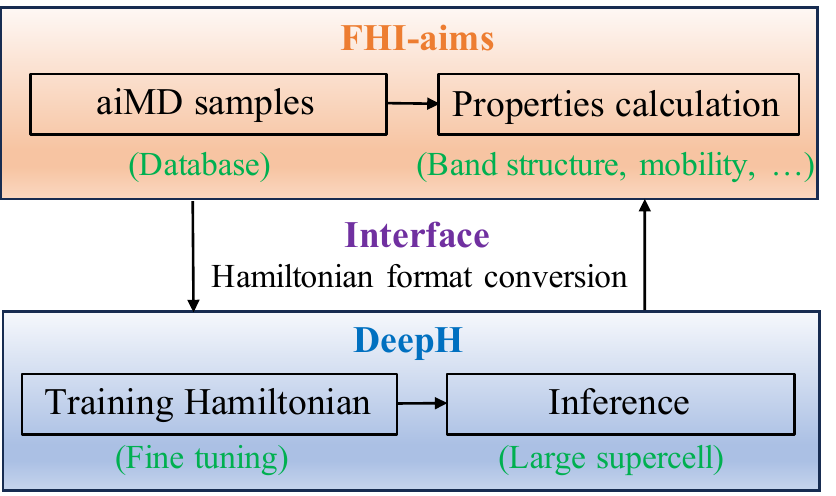}
\caption{The workflow for large supercell Hamiltonian prediction using \texttt{DeepH} and subsequent physical properties calculation implemented in \texttt{FHI-aims}. }
\label{fig:workflow}
\end{figure}

\texttt{FHI-aims} and \texttt{DeepH} are integrated to enable efficient calculations of electronic properties within the proposed AI-assisted framework, as illustrated in Fig.\,~\ref{fig:workflow}.
Training data are generated from \textit{ab initio} MD\,(aiMD) simulations performed in \texttt{FHI-aims}, which provide atomic structures together with corresponding Kohn--Sham Hamiltonian matrix elements in a localized atomic orbital basis obtained from fully converged SCF calculations.
The Atomic Simulation Interface\,(\texttt{ASI}) toolkit\,\cite{2023atomic} is then used to convert the \texttt{FHI-aims} outputs into graph-based datasets suitable for \texttt{DeepH} training.
Using these datasets, the neural network is trained to learn the mapping between atomic configurations and Hamiltonian matrices.
To improve the model's transferability, additional samples with larger supercells and different temperatures are included during fine-tuning.

For new atomic configurations, \texttt{DeepH} directly predicts the corresponding Hamiltonian without performing DFT calculations in a scf loop.
The predicted Hamiltonian can then be diagonalized to obtain electronic properties such as band structures and densities of states\,(DOS).
In the present work, the predicted Hamiltonian is converted back into the \texttt{FHI-aims} format to enable further calculations of transport-related quantities via the aiKG framework.
In particular, we focus on electronic transport properties, particularly carrier mobility, as well as the computational acceleration enabled by the predicted Hamiltonian.

\subsection{Kubo--Greenwood Formalism}
Electronic transport properties are evaluated using the aiKG formalism implemented in \texttt{FHI-aims}.
Within this framework, the frequency-dependent electrical conductivity is expressed in terms of the current-current correlation function\,\cite{kubo1957,holst2011},
\begin{equation}
\sigma_{i j}(\omega)=\lim _{\alpha \rightarrow 0^{+}} \int_0^{\infty} d t e^{i(\omega+i \alpha) t} \int_0^\beta d \tau \operatorname{Tr}\left\langle \hat{\mathbf{J}}_i(t-i \hbar \tau) \cdot \hat{\mathbf{J}}_j\right\rangle
\end{equation}
where $\beta$ = (k$_B$T)$^{-1}$ denotes the inverse thermal energy, and $\alpha$ is an infinitesimal adiabatic parameter introduced to ensure convergence of the response function.
In the effective single-particle approximation, the current operator is given by\,\cite{greenwood1958,holst2011}
\begin{equation}
\hat{\mathbf{J}}=\frac{q}{m} \sum_{\mathbf{k} \mu \nu}\langle\mathbf{k} \mu| \hat{\mathbf{p}}|\mathbf{k} \nu\rangle|\mathbf{k} \mu\rangle\langle\mathbf{k} \nu|,
\end{equation}
where $q$ and $m$ are the charge and the mass of the carrier, respectively. $| \mathbf{k}\mu\rangle$ denotes the Bloch state with crystal momentum $\mathbf{k}$ and band index $\mu$, and $\hat{\mathbf{p}}$ represents the momentum operator.
The Kubo--Greenwood formalism thus provides a quantum-mechanical description of electrical conductivity based on the electronic structure and current matrix elements.
More generally, the Kubo formalism offers a unified framework for describing diverse transport mechanisms, including electronic, polaronic\,\cite{2022optical}, and ionic conductivity\,\cite{gigli2024}, provided that the corresponding current operators are appropriately defined.

The real part of the KG formula is used to evaluate the frequency-dependent electrical conductivity arising from interband transitions at the same crystal momentum\,\cite{holst2011},
\begin{equation}
{\begin{aligned}
\Re\left(\sigma_{i i}(\omega)\right)= & \frac{2 \pi q^2 \hbar^2}{m^2 V \omega} \sum_{\mathbf{k} \mu \nu}\left(\langle\mathbf{k} \nu| \nabla_i|\mathbf{k} \mu\rangle\langle\mathbf{k} \mu| \nabla_i|\mathbf{k} \nu\rangle\right) \\
& \times\left(f_{\mathbf{k} \nu}-f_{\mathbf{k} \mu}\right) \delta\left(\epsilon_{\mathbf{k} \mu}-\epsilon_{\mathbf{k} \nu}-\hbar \omega\right),
\end{aligned}}
\label{eq:conductivity}
\end{equation}
where V is the volume of the supercell. $f_{\mathbf{k}\mu}$ and $\epsilon_{\mathbf{k}\mu}$ denote the occupation numbers and eigenvalues of the electronic states, respectively.
The diagonal components $\sigma_{ii}$ correspond to the longitudinal conductivity. Only transitions between electronic states $|\mathbf{k}\mu\rangle$ and $|\mathbf{k}\nu\rangle$ with the same crystal momentum $\mathbf{k}$ are considered.
The corresponding imaginary part can then be obtained from the real part of the conductivity through the Kramers--Kronig relation\,\cite{kubo1957}.

In the all-electron framework of \texttt{FHI-aims}, which employs localized real-space numeric atomic orbitals\,(NAOs), the momentum matrix elements are expressed in terms of the real-space basis functions $\phi_{j\mathbf{N}}(r)$,
\begin{equation}
\begin{gathered}
\langle\mathbf{k} \mu| \nabla|\mathbf{k} \nu\rangle=\sum_{i j}\left[C_\mu^i(\mathbf{k})\right]^* C_\nu^j(\mathbf{k}) \sum_{\mathbf{N}} e^{i \mathbf{k} \cdot \mathbf{T}(\mathbf{N})} \\
\times\left\langle\phi_{i \mathbf{0}}(r)\right| \nabla\left|\phi_{j \mathbf{N}}(r)\right\rangle ,
\end{gathered}
\end{equation}
where $C_\mu^i(\mathbf{k})$ is the expansion coefficient of the Bloch eigenstate $| \mathbf{k}\mu\rangle$. $\phi_{j \mathbf{N}}(r)$ denotes the basis function $j$ located in the unit cell $\mathbf{N}$, $\mathbf{T}(\mathbf{N})$ is the lattice translation vector, and $\langle\phi_{i \mathbf{0}}(r)|\nabla|\phi_{j \mathbf{N}}\rangle$ represents the gradient matrix elements of the NAO basis functions\,\cite{knuth2015,shang2017}.
The electronic states are obtained by solving the generalized Kohn--Sham equation
\begin{equation}
\mathbf{H}\left(\mathbf{k}\right)\left|\mathbf{k} \mu\right\rangle=\epsilon_{\mathbf{k}\mu} \mathbf{S}\left(\mathbf{k}\right)\left|\mathbf{k} \mu\right\rangle,
\end{equation}
where $\mathbf{H}$ and $\mathbf{S}$ denote the Hamiltonian and overlap matrices in reciprocal space.
To efficiently evaluate the conductivity on dense $\mathbf{k}$ grids, Fourier interpolation of the KG quantities is employed in \texttt{FHI-aims}.

The DC limit of the electrical conductivity at finite temperatures is calculated using sufficiently large supercells combined with long MD simulations.
Large supercells mitigate two key limitations of finite-size calculations.
First, they induce band folding, which converts intraband transitions in the primitive cell into interband transitions in the supercell and thereby recovers contributions that are otherwise missing in the DC limit.
Second, larger supercells allow the inclusion of long-wavelength lattice vibrations associated with increasingly small energy differences $\hbar\omega$ in Eq.\,\ref{eq:conductivity}, which are essential for reliability in describing finite-temperature transport.
In addition, long MD simulations with sufficient statistical sampling are required to capture thermal fluctuations, dynamical disorder, and anharmonic lattice effects.
To adequately sample the accessible phase space of the canonical ensemble, transport properties are therefore evaluated as ensemble averages over multiple thermally generated atomic configurations.

\subsection{Spectral Function}
Electronic spectral functions are constructed using a band-unfolding approach\,\cite{quan2025} that maps the folded electronic band structure of a supercell onto the primitive-cell Brillouin zone.
This procedure enables the extraction of spectral information from supercell configurations generated by aiMD simulations.
The spectral function is defined as the imaginary part of Green's function\,\cite{bruus2004,coleman2015},
\begin{equation}
A(\mathbf{k}, E)=\sum_{K N} W_{\mathbf{K} N}^{\mathbf{k}} \delta\left(E-E_{\mathbf{K} N}\right),
\end{equation}
where $E_{\mathbf{K}N}$ stands for the eigenvalue of the supercell electronic state $|\Psi_{\mathbf{K}N}\rangle$.
The unfolding weight is given by $W_{\mathbf{K} N}^{\mathbf{k}}=\sum_n\left|\left\langle\psi_{\mathbf{k} n} \mid \Psi_{\mathbf{K} N}\right\rangle\right|^2 $, where $|\psi_{\mathbf{k}n}\rangle$ and $|\Psi_{\mathbf{K}N}\rangle$ represent the electronic states of the primitive cell and the supercell, respectively.

In the non-orthogonal linear-combination-of-atomic-orbitals\,(LCAO) basis utilized in \texttt{FHI-aims}, the unfolding weights can be expressed as\,\cite{quan2025}
\begin{equation}
W_{\mathbf{K} N}^{\mathbf{k}}=\sum_n\left|\mathbf{F}_{\mathbf{k} n}^{\dagger} \mathbf{C}_{\mathbf{K} N}^{\prime}\right|^2, 
\end{equation}
where $\mathbf{F}_{\mathbf{k} n}$ denotes the eigenvector of the primitive-cell Hamiltonian. The $\mathbf{C}_{\mathbf{K} N}$ is the supercell wave function.
The temperature-dependent spectral function\cite{zacharias2020,quan2025} is obtained as a canonical ensemble average over atomic configurations sampled from MD simulations,
\begin{equation}
\langle A(\mathbf{k}, E)\rangle_T=\lim _{I \rightarrow \infty} \sum_i^I A^i(\mathbf{k}, E),
\end{equation}
where the index $i$ labels a nuclear configuration corresponding to temperature $T$.
Within perturbation theory, the spectral function typically exhibits a Lorentzian line shape that reflects well-defined quasiparticle states\,\cite{giustino2017}.
In contrast, the present approach evaluates spectral functions by averaging over thermally sampled MD configurations, thereby incorporating anharmonic lattice dynamics and electron-vibration coupling to all orders.
As a result, the spectral functions can exhibit asymmetric, non-Lorentzian line shapes, particularly at elevated temperatures\cite{quan2025}.

\section{Results and Discussion}

\begin{figure*}[ht!]
\centering
\includegraphics[width=1\linewidth]{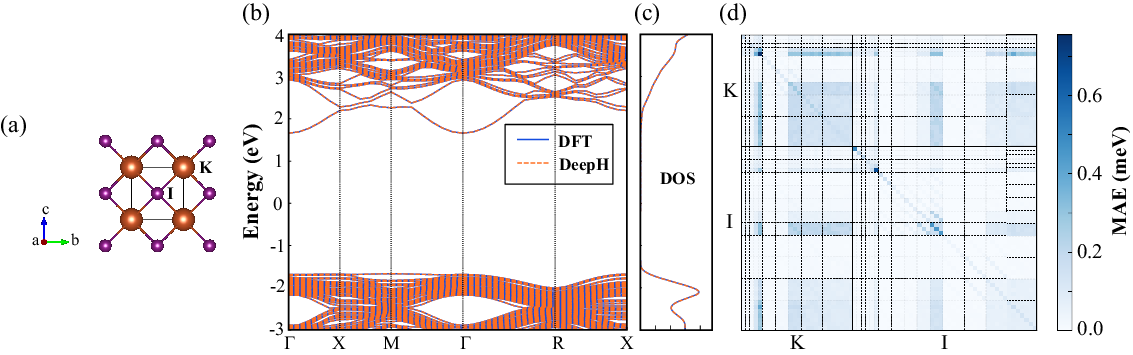}
\caption{ (a) Crystal structure of the primitive cell KI. Comparison of electronic structures calculated using DFT and \texttt{DeepH} Hamiltonians of 3 $\times$ 3 $\times$ 3 supercell KI with 54 atoms in (b) electronic band structure and (c) DOS. (d) Mean absolute error (MAE) of the predicted Hamiltonian for K–I orbital interactions, evaluated with basis sets \textit{s5p3d1f1} for K and \textit{s6p5d3f1} for I.}
\label{fig:bandstructure}
\end{figure*}

\subsection{Strongly Anharmonic Material KI}
\label{subsec:KI}
Potassium iodide\,(KI) is chosen as an example for strongly anharmonic material to benchmark the present methodology, which has been experimentally shown to possess strong temperature-dependent ionic displacements\cite{KI1967}.
KI crystallizes in a cubic ionic structure belonging to the Pm$\bar{3}$m\,(\#221) space group, as shown in Fig.\,\ref{fig:bandstructure} (a), and was identified as a promising candidate for strong lattice anharmonicity in a high-throughput screening of more than 700 materials via a symbolic regression model implemented by Sure Independence Screening and Sparsifying Operator\,(SISSO)\,\cite{purcell}.
According to the data-driven screening framework proposed by Purcell \textit{et al.}\,\cite{purcell}, lattice thermal transport is largely governed by three descriptors: strong anharmonicity, a low high-temperature Debye temperature, and a large molar volume.
KI simultaneously satisfies these criteria.
In particular, it exhibits an anharmonicity parameter of $\sigma_{A} = 0.4$ at room temperature, well above the commonly adopted threshold $\sigma_{A} > 0.2$ for strong higher-order lattice interactions.
In addition, its relatively low Debye temperature (131\,K) and large molar volume (77.6\,$\text{\AA}^3$) indicate a soft phonon spectrum and enhanced vibrational phase space for phonon scattering.
These characteristics make KI a particularly demanding system for theoretical approaches based on perturbative or quasi-harmonic descriptions.

The relatively flat valence band of KI further makes it suitable for transport analysis within the KG framework, as the flat band ensures that the interband transition contributions converge rapidly compared with dispersive band\,\cite{KG}.
In the KG formalism implemented in \texttt{FHI-aims}, electrical conductivity is evaluated from interband electronic transitions at identical crystal momentum.
As a consequence, intraband contributions that would normally appear in the DC limit are not directly captured in primitive-cell calculations with interband transitions.
Increasing the supercell size introduces band folding in reciprocal space, thereby converting some intraband transitions in the primitive cell into interband transitions in the supercell.
In the limit of an infinitely large supercell, both intra- and interband contributions would therefore be recovered.
In practice, however, such calculations are computationally prohibitive, and an appropriate supercell size must be chosen to obtain converged transport properties.
For electronic bands with relatively small dispersion, band folding does not significantly alter the electronic structure and thus enables faster convergence with respect to supercell size.
In KI, the valence band is considerably flatter than the conduction band, and we therefore focus on hole transport associated with the valence-band maximum in the following analysis.


\subsection{Electronic Band Structure}
\label{subsec:band}
Before assessing transport properties, we first examine the electronic structure of KI and the reliability of the \texttt{DeepH}-predicted Hamiltonian.
Electronic band structure and species-projected DOS of KI from DFT primitive-cell calculations are shown in Fig.\,S1 in the Supplementary Material\,(SM).
KI  exhibits an indirect band gap of 3.375\,eV within the DFT-PBE formalism, with the conduction-band minimum\,(CBM) located at the $\mathrm{M}$ point and the valence-band maximum\,(VBM) at the $\Gamma$ point.
The calculated hole effective mass at the VBM is 2.56\,$m_{e}$.
The projected DOS indicates that K orbitals primarily contribute to the conduction band, while I orbitals dominate the valence band, reflecting the strongly ionic nature of KI. The experimental band gap of KI is approximately 6\,eV\,\cite{KI-experiment1,KI-experiment2}, which is larger than the calculated PBE value.
This deviation minimally affects the  mobility, as it depends primarily on band dispersion and scattering phase space, not the absolute gap. In addition, our work focuses on benchmarking deep-learning against explicit DFT rather than matching experiment, this functional-induced variation is negligible for our conclusions.

Fig.\,\ref{fig:bandstructure} (b, c) compares the electronic band structures and DOS calculated using full DFT and the \texttt{DeepH}-predicted Hamiltonian for a $3\times3\times3$ KI supercell (54 atoms).
The predicted eigenvalues agree well with the DFT results, demonstrating the high accuracy in training of the \texttt{DeepH} model, with a mean absolute error\,(MAE) of approximately 5.6\,meV.
To quantify the prediction reliability, we analyze the errors in the Hamiltonian matrix elements.
As shown in Fig.\,\ref{fig:bandstructure} (d), the MAE associated with individual K-I orbital interactions ($69\times69$ orbitals) remains below 0.8\,meV, while the diagonal elements exhibit comparatively larger deviations, indicating that the dominant contribution to the overall error arises from the on-site (self-atomic) energies.
The average MAE of the full Hamiltonian matrix is approximately 0.05\,meV for the training set and 0.10\,meV for the test set.
Importantly, the model trained on the 54-atom supercell and fine-tuned with a small number of additional configurations generalizes well to larger supercells containing 128 and 250 atoms, the Hamiltonian MAE evaluated on the test data remains as low as 0.17 and 0.36\,meV, respectively, substantiating the good scalability and transferability of the trained model.
However, accurate predictions for substantially larger supercells (e.g., >\,500 atoms) remain challenging due to the long-range electrostatic interactions inherent to ionic crystals, which will be the focus of our future work.

\begin{figure}[ht!]
\includegraphics[width=1\linewidth]{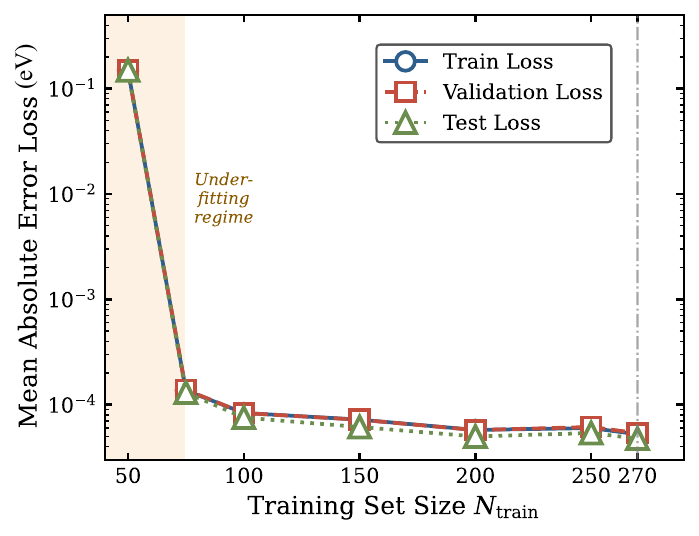}
\caption{Convergence of the DeepH training, validation and test losses as a function of training-set size $N_{\mathrm{train}}$ of 54-atom supercell. MAE losses (eV) are plotted on a logarithmic scale for six partitions of the KI configuration pool, with validation and test sets kept roughly proportional to $N_{\mathrm{train}}$.}
\label{fig:convergence}
\end{figure}

To verify that the accuracy originates from a genuinely converged model rather than a specific choice of training subset, we systematically varied the training-set size from $N_{\mathrm{train}} = 50$ to $N_{\mathrm{train}} = 270$ with proportionally scaled validation and test partitions (Fig.~\ref{fig:convergence}).
When $N_{\mathrm{train}} = 50$ the loss is near $1.5 \times 10^{-1}$\,eV, indicating that the network has not yet learned an accurate mapping for Hamiltonian.
Between $N_{\mathrm{train}} = 50$ and $N_{\mathrm{train}} = 200$ the learning threshold is crossed and the loss drops by three orders of magnitude.
The incremental improvement also becomes progressively smaller.
The reduction in loss between$N_{\mathrm{train}} = 250$ and $N_{\mathrm{train}} = 270$ is already below $13\,\%$.
At the adopted $N_{\mathrm{train}} = 270$, the train, validation and test losses reach $5.25 \times 10^{-5}$, $5.38 \times 10^{-5}$ and $4.77 \times 10^{-5}$\,eV, respectively, with a relative generalization gap below $3\,\%$ and the test loss consistently falling below the training loss across the entire converged regime, indicating good generalization without evidence of overfitting.
Given the absence of a diverging train-validation gap and the consistency with aiMD sampling, a training set of 270 configurations is deemed sufficient to capture the accessible thermal fluctuations in strongly anharmonic KI. The active-learning strategy will be discussed in more detail below.

\begin{figure}[ht!]
\includegraphics[width=0.9\linewidth]{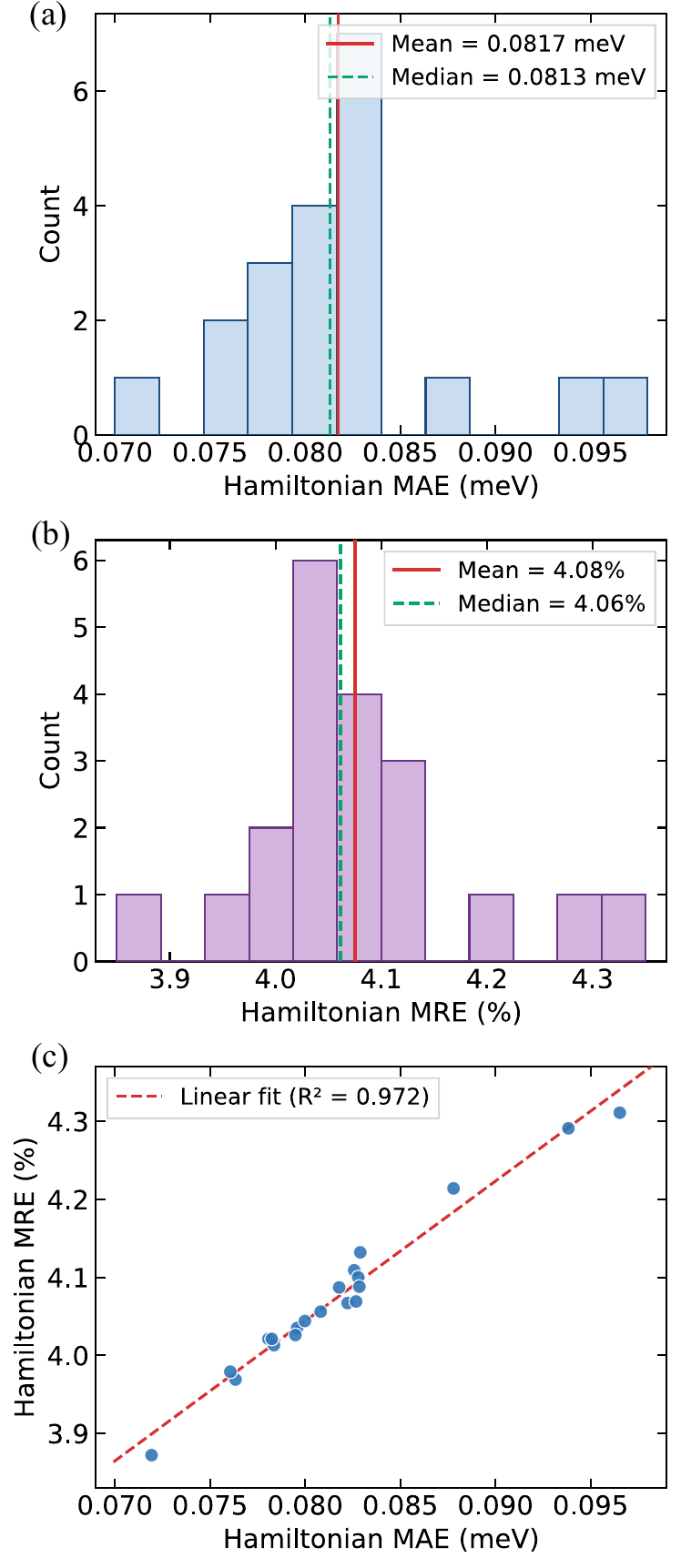}
\caption{Statistical analysis of the per-configuration Hamiltonian prediction errors for the \texttt{DeepH} model evaluated on 20 held-out test configurations of the 54-atom KI supercell at 300 K. (a) Distribution of the mean absolute error (MAE) of the predicted Hamiltonian matrix elements. (b) Distribution of the mean relative error (MRE). In (a) and (b), vertical lines mark the mean (red solid) and median (green dashed). (c) Correlation between MAE and MRE for each test configuration, with a linear fit (R$^{2}$ = 0.972) shown as a red dashed line.}
\label{fig:hamiltoian}
\end{figure}

Beyond the average error metrics, we examine the statistical distribution of Hamiltonian prediction errors across individual test configurations to assess the robustness and uniformity of the trained model\,(Fig.\,\ref{fig:hamiltoian}).
The per-configuration MAE evaluated over 20 held-out test configurations exhibits a narrow, approximately symmetric distribution with a mean of 0.0817 meV and a standard deviation of only 0.006 meV.
The ratio of the maximum MAE\,(0.097 meV) to the mean is 1.18, confirming the absence of heavy-tailed outliers.
The mean relative error\,(MRE) follows a similarly compact distribution centered at 4.08\,\%, with mean and median values nearly coincident\,(4.08\,\% and 4.06\,\%), further indicating that no subset of configurations is systematically harder to predict.
Moreover, the MAE and MRE across configurations are strongly linearly correlated\,(R$^2$ = 0.972), indicating that the prediction errors are governed by a single, uniform mechanism rather than by configuration-dependent failure modes.
These results provide direct evidence that the aiMD training protocol generates a well-covered training distribution, which the model does not extrapolate for any of the test configurations, and no spurious rare events are present in the sampled configuration space.
This uniformity is a prerequisite for the reliable evaluation of transport properties in the subsequent sections, where even small systematic errors in the Hamiltonian can be amplified through the KG formalism.

The computational cost of generating training data could potentially be reduced by replacing aiMD with molecular dynamics driven by machine-learned interatomic potentials (MLIP-MD), which enables much more efficient configurational sampling\,\cite{2019machine,lorenz2004} However, because MLIP-MD samples an approximate potential, it may fail to capture physically important configurations in strongly anharmonic materials\,\cite{Kang2025}. Although not considered in the present work, integrating active-learning workflows such as \texttt{ALmoMD}\cite{Kang2025} with the DeepH data-generation pipeline could provide an efficient route toward aiMD-quality sampling at substantially lower computational cost.

\subsection{Transport properties}
\label{subsec:mobility}
Carrier transport in strongly anharmonic materials is investigated here within the KG formalism, which naturally incorporates full anharmonic lattice dynamics and electron-vibration interactions.
We assess our DeepH-aiKG workflow against its underlying DFT-aiKG reference, not directly against experiment. This reference has itself been validated against measured transport in strongly anharmonic SrTiO$_3$ and BaTiO$_3$\cite{KG}.
In this framework, the electronic band structure largely determines the strength of electron-vibration coupling and therefore plays a central role in limiting carrier transport.
In KI, the valence band is significantly flatter than the conduction band, allowing long-wavelength contributions to electron-vibration coupling to be captured more efficiently in finite supercells and facilitating faster convergence with respect to supercell size.
In both experimental and theoretical studies, carrier mobility is often preferred over electrical conductivity because it more directly reflects intrinsic transport properties, is independent of carrier concentration, and is less sensitive to the band gap\,\cite{zhou2018,ponce2018}.
We therefore focus on the hole mobility of KI in the following analysis, where the mobility is defined as $\mu = \sigma/(qn)$, with the electrical conductivity $\sigma$, the elementary charge $q$, and the carrier concentration $n$.

\begin{figure}[t!]
\centering
\includegraphics[width=1\linewidth]{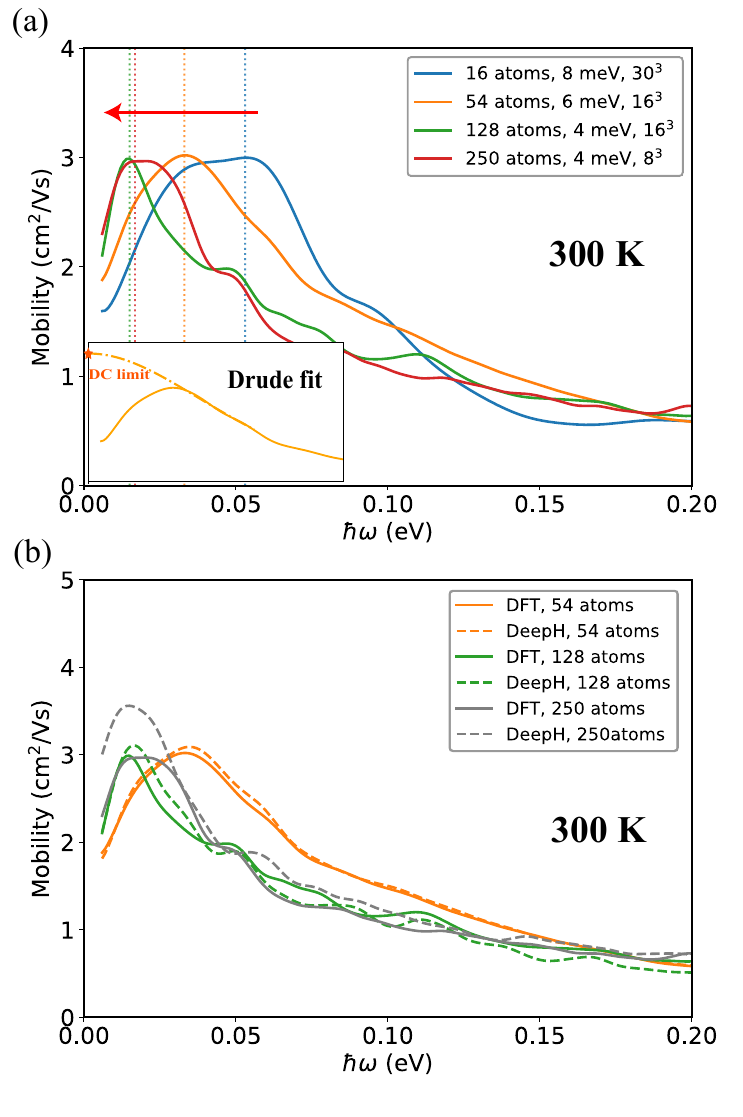}
\caption{(a) Supercell size dependent hole mobility at 300 K with adjustable Gaussian broadening parameter $\eta$ and the density of the $\mathbf{k}$-point grid. Supercell 16 atoms: $2\times2\times2$, 54 atoms: $3\times3\times3$, 128 atoms: $4\times4\times4$, 250 atoms: $5\times5\times5$. The dashed lines are peak positions of the mobility spectral. The inset is the illustration of the Drude fit for the 54-atom KI supercell at 300\,K. (b) The comparison of DFT and \texttt{DeepH} Hamiltonian mobility calculation at 300 K with different supercell sizes (54 atoms, 128 atoms and 250 atoms) of KI. The model was pre-trained on 270 configurations of 54-atom KI and subsequently fine-tuned using a small number configurations of 128-atom KI.}
\label{fig:mob-1}
\end{figure}

A correct description of the DC mobility requires approaching the low-frequency limit ($\omega \rightarrow 0$).
Capturing these long-wavelength contributions directly would require extremely large supercells and is therefore computationally prohibitive.
For this purpose, we follow the approach employed for anharmonic SrTiO$_3$\cite{KG}, i.e. the DC limit is obtained by extrapolating the low-frequency mobility spectrum using a Drude model equation\,\cite{drude1900,willis2013}.
In the low-frequency regime, the mobility can be approximated as
\begin{equation}
\mu(\omega) \approx \frac{\mu_0}{(\omega \tau)^2+1},
\end{equation}
where $\mu_0$ denotes the DC-limit mobility and $\tau$ represents the effective carrier lifetime.
As an illustrative example, the Drude fit for the 54-atom KI supercell at 300\,K is shown in the inset of Fig.~\ref{fig:mob-1} (a).
Although the Drude model provides a practical approach for DC extrapolation, it may fail for non-Drude-like low frequency responses. For example, in nondegenerate plasmas, the proportionality of the relaxation time generates significantly different frequency dependencies\,\cite{Redmer}.

\begin{figure}[ht!]
\centering
\includegraphics[width=0.9\linewidth]{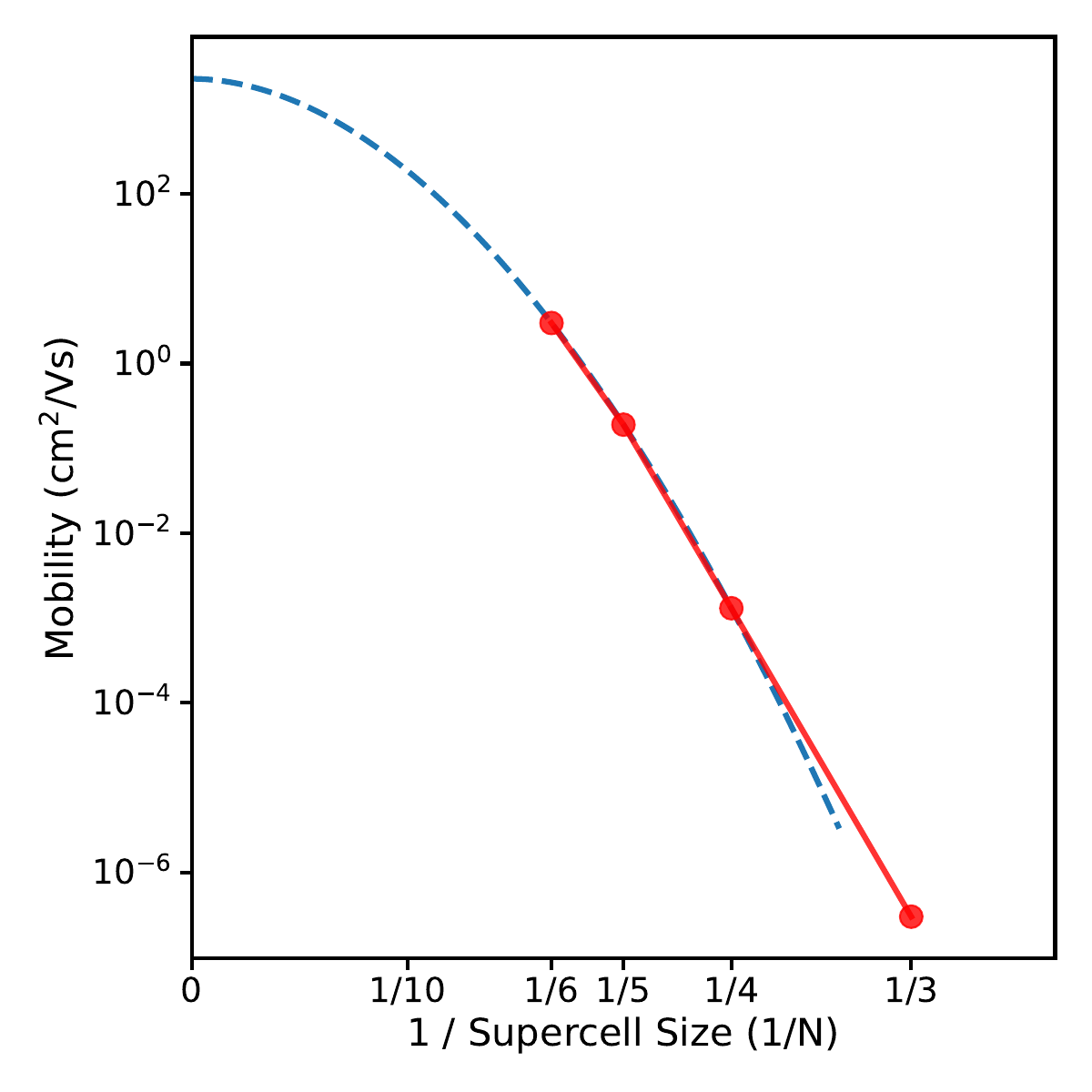}
\caption{ Supercell size dependent electron mobility at 300\,K of KI. N is the supercell linear dimension. The red circles represent the calculated mobility values for different supercell sizes, 54 atoms: $3\times3\times3$, 128 atoms: $4\times4\times4$, 250 atoms: $5\times5\times5$, 432 atoms: $6\times6\times6$. The dashed curve is an empirical extrapolation fitted to the data, providing an extrapolation of the mobility toward the infinite supercell limit (1/N$\rightarrow$0).}
\label{fig:ele_mobility}
\end{figure}

The pronounced size dependence in Fig.\,\ref{fig:ele_mobility} spanning over six orders of magnitude from the $3\times3\times3$ to the
$6\times6\times6$ supercell, is a direct manifestation of the highly
dispersive KI conduction band. In the primitive cell, the DC mobility
is dominated by intraband contributions, which are absent from the
interband-only KG expression in Eq.\,\ref{eq:conductivity}; band folding upon
supercell construction is precisely the mechanism by which these missing
intraband contributions are progressively recovered. For a flat band,
such as the KI valence band, a modest supercell already folds the
relevant phase space onto the same crystal momentum and convergence is
rapid. For a dispersive band, by contrast, the supercell required to
fold the relevant intraband transitions onto $\omega \rightarrow 0$ grows
rapidly, so that charge carriers remain artificially confined and the
computed mobility is strongly suppressed at accessible supercell sizes.
The mobility is consequently far from converged across the range studied,
and reaching the thermodynamic limit by brute-force supercell enlargement
is computationally prohibitive for the conduction band of this system.

This observation carries a general implication for non-perturbative
transport calculations. Any DC mobility extracted for a dispersive band
on a supercell smaller than the convergence regime should be regarded as
a lower bound rather than a converged estimate, since the intraband
contributions recovered by band folding remain incompletely captured.
We expect this caveat to apply broadly to ionic materials that combine a
flat band of one carrier type with a dispersive band of the other, for
which the two carrier channels converge at markedly different supercell
sizes. The flat KI valence band, by contrast, permits well-converged
hole mobilities at the comparatively small supercells used in the
remainder of this work, which is why the subsequent transport analysis
focuses on hole transport.

We stress that the stretched-exponential curve in Fig.\,\ref{fig:ele_mobility} is an empirical extrapolation rather than a physically derived functional
form. It is adopted because the data follow a smooth, monotonic trend
over six orders of magnitude and a stretched exponential captures both
the steep rise across the accessible range and the saturation expected
as 1/N$\rightarrow$0. Only the three largest supercells were included in the fit, as the smallest supercell is expected to exhibit stronger finite-size effects and therefore does not reliably represent the scaling behavior. The extrapolated value of order
$10^{3}$~cm$^2$V$^{-1}$s$^{-1}$ should accordingly be read as an
order-of-magnitude estimate of the thermodynamic-limit electron mobility,
not as a precise prediction.

The convergence of the hole mobility with respect to supercell size is shown in Fig.\,\ref{fig:mob-1} (a), obtained from an average over 50 independent snapshots from MD simulations at 300\,K.
As the supercell size increases, the peak of the frequency-dependent mobility spectrum gradually shifts toward the DC limit, indicating improved sampling of long-wavelength contributions to electron-vibration scattering.
In these calculations, both the Gaussian broadening parameter $\eta$ and the density of the $\mathbf{k}$-point grid are adjusted consistently with the supercell size to ensure numerical convergence. 
After applying the Drude extrapolation, the DC-limit hole mobilities obtained from the DFT Hamiltonian are 3.76 and 1.82\,cm$^{2}$V$^{-1}$s$^{-1}$ at 300 and 900\,K, respectively.
To further verify the convergence behavior, mobilities calculated for larger supercells at 300\,K are also examined.
For the 128-atom and 250-atom supercells, the DFT-calculated mobilities at 300\,K are 3.48 and 3.60 cm$^{2}$V$^{-1}$s$^{-1}$, respectively, in good agreement with the results obtained for the 54-atom supercell. Owing to the relatively flat valence band of KI, convergence with respect to supercell size can be achieved using comparatively small supercells, with the DC mobility varying by only $\sim\!7\,\%$ and $\sim\!3\,\%$ between consecutive supercell sizes.

We next examine the capability of the \texttt{DeepH}-predicted Hamiltonian to reproduce the carrier mobility spectrum.
Fig.\,\ref{fig:mob-1} (b) compares the hole mobility obtained from the full DFT Hamiltonian with that derived from the \texttt{DeepH}-predicted Hamiltonian for 54-, 128- and 250-atom KI supercells at 300\,K.
The overall spectrum shape and peak positions are well reproduced, demonstrating the strong predictive performance of \texttt{DeepH} for transport properties.
For the larger 128-atom supercell, MAE of the band structure increases to $\sim$8.2\,meV, compared with $\sim$5.6\,meV for the 54-atom case.
Although this increase remains within an acceptable range, the mobility spectrum is sensitive to the underlying electronic structure and typically requires eigenvalue errors below approximately 10\,meV.
Consequently, the deviation between the \texttt{DeepH}-predicted and DFT mobilities becomes slightly larger for the 128-atom supercell than for the 54-atom case, and this discrepancy grows for the 250-atom supercell.

\begin{figure}[ht!]
\centering
\includegraphics[width=1\linewidth]{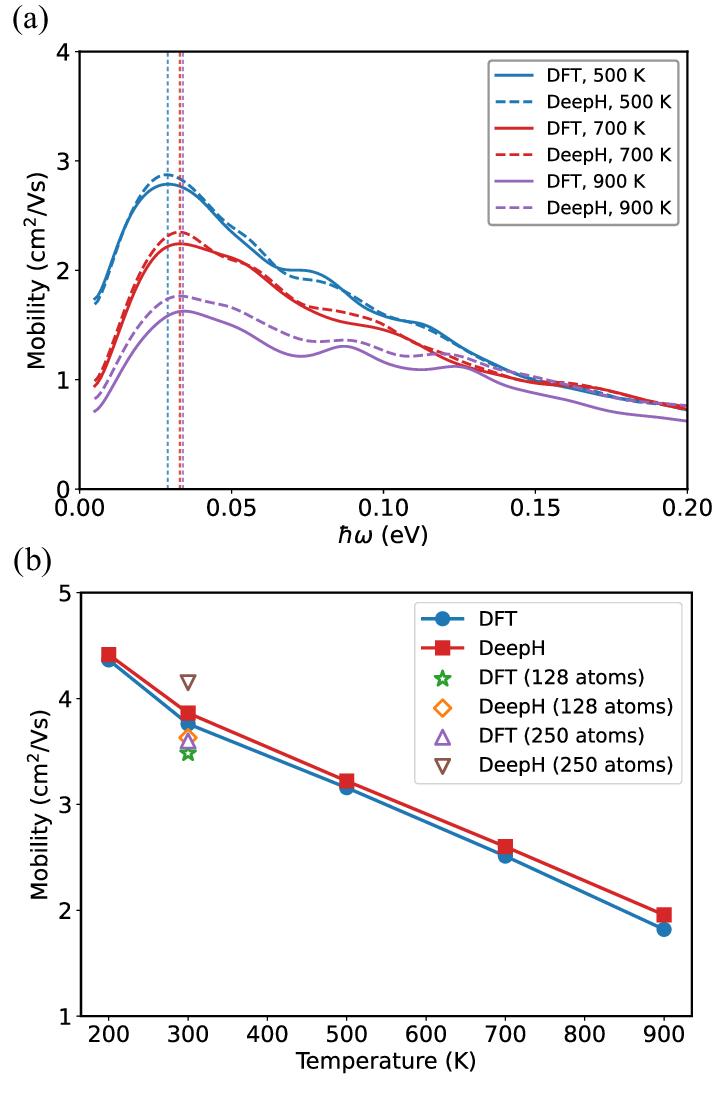}
\caption{(a) Frequency-dependent mobility comparing of DFT and \texttt{DeepH} Hamiltonians at several high temperatures. (b) Temperature-dependent predicted mobility of KI with Drude fit. Four additional data points for supercells of 128 and 250 atoms are shown for comparison. The model was pre-trained on 270 configurations of 54-atom KI and subsequently fine-tuned using a small number of 128-atom KI configurations, as well as 54-atom KI configurations at various temperatures.}
\label{fig:mob-2}
\end{figure}

The temperature dependence of the carrier mobility is shown in Fig.\,\ref{fig:mob-2} (a) for the 54-atom KI supercell.
The calculated mobility decreases systematically with increasing temperature.
At elevated temperatures, enhanced lattice vibrations increase electron-vibration scattering, thereby shortening carrier lifetimes and reducing mobility.
This behavior is consistent with the well-established dominance of lattice-scattering mechanisms in semiconductors at high temperatures, as widely reported in the literature\,\cite{giustino,ponce2020,li2015}.

The temperature-dependent mobility spectra obtained using the \texttt{DeepH}-predicted Hamiltonian closely reproduce the DFT results, as shown in Fig.\,\ref{fig:mob-2} (a).
After applying the Drude extrapolation, the DC-limit hole mobilities predicted by \texttt{DeepH} are 3.86 and 1.96\,cm$^{2}$V$^{-1}$s$^{-1}$ at 300 and 900\,K, respectively.
The temperature dependence follows a power-law behavior, with exponents of $T^{-0.487}$ for the DFT results and $T^{-0.464}$ for the \texttt{DeepH} predictions, indicating very similar transport trends.
The mobility exponent is markedly weaker than the $T^{-3/2}$ dependence predicted by deformation-potential theory for
transport limited by acoustic-phonon scattering~\cite{bardeen1950}, and weaker still than the steeper temperature dependence characteristic of polar-optical (Fr\"ohlich) coupling~\cite{frohlich1954,giustino2017}.
Such a weak dependence is consistent with the non-perturbative, strongly anharmonic regime probed here, in which acoustic, optical, and higher-order anharmonic scattering channels contribute simultaneously~\cite{xia2020} and the carrier scattering approaches the Ioffe--Regel limit~\cite{ioffe}, rather than with a single dominant mechanism whose canonical exponent would be recovered.
To further assess the robustness of the predictions, mobilities for larger supercells at 300\,K are also examined.
For the 128-atom and 250-atom supercells, the \texttt{DeepH}-predicted mobilities are 3.63 and 4.15\,cm$^{2}$V$^{-1}$s$^{-1}$, respectively.
As supercell sizes increase, accumulated numerical uncertainties associated with KG interpolation and low-frequency extrapolation may become non-negligible, underscoring the need for careful validation when applying \texttt{DeepH} to extended systems.
The DeepH-DFT mobility deviation grows monotonically and faster than linearly with the Hamiltonian MAE, rising from 2.7\,\% to 4.3\,\% and
15.3\,\% as the MAE increases from 0.10 to 0.17 and 0.36\,meV.
Although additional datasets are used to fine-tune the model at elevated temperatures, the deviations between the DFT and predicted Hamiltonians remain slightly larger than those at lower temperatures, reflecting the increased configurational complexity and expanded phase space of atomic environments at high temperatures.
To address this, future work will focus on deliberately constructing training datasets that encompass larger atomic displacements, particularly under high-temperature conditions, to enhance the model's generalizability and robustness.


\subsection{Temperature-dependent spectral function and effective mass}
\label{subsec:spectral}

\begin{figure}[t!]
\centering
\includegraphics[width=1\linewidth]{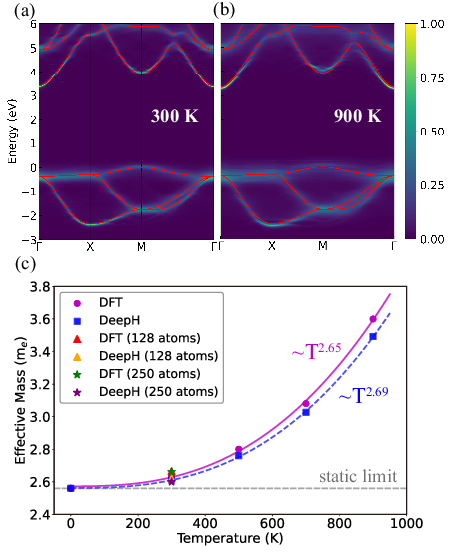}
\caption{Spectral function obtained with the DFT Hamiltonian at (a) 300 K and (b) 900 K of 54 atoms KI. (c) The temperature-dependent effective mass of KI at VBM (M point) with DFT and \texttt{DeepH} Hamiltonian. Four additional data points for supercells of 128 and 250 atoms are also shown for comparison. The gray dashed line denotes the effective mass result in the static limit. The model was pre-trained on 270 configurations of 54-atom KI and subsequently fine-tuned using a small number of 128-atom KI configurations, as well as 54-atom KI configurations at various temperatures.}
\label{fig:spectral}
\end{figure}

The spectral functions of KI exhibit pronounced temperature-induced broadening and band-gap renormalization.
The fully anharmonic and thermodynamically averaged spectral functions obtained from the band-unfolding approach are shown in Fig.~\ref{fig:spectral} (a, b) for temperatures of 300 K and 900 K.
As the temperature grows, the spectral features broaden significantly.
The linewidth of the spectral function is directly related to the imaginary part of the electron self-energy, reflecting enhanced electron–vibration coupling and a corresponding reduction in carrier lifetime.
At the same time, the shift of the spectral-function peak position is governed by the real part of the electron self-energy, which describes the temperature-dependent quasiparticle energy renormalization.
As a result, the band gap between the VBM at the $\Gamma$ point and the CBM at the $\mathrm{M}$ point decreases from 3.35 eV at 300 K to 3.25 eV at 900 K.

\texttt{DeepH} reliably reproduces the temperature-dependent spectral functions obtained from DFT.
The spectral functions derived from the \texttt{DeepH}-predicted Hamiltonian are shown in SM\,(Fig.\,S2).
As discussed previously for the mobility calculations, the expanded phase space of atomic configurations at elevated temperatures introduces greater structural diversity in the local atomic environments, making the learning task more challenging.
Nevertheless, the \texttt{DeepH} results remain in strong agreement with the DFT spectral functions up to 900\,K, demonstrating the robustness and transferability of the \texttt{DeepH} framework under strongly anharmonic and high-temperature conditions.

The spectral functions also enable the extraction of the temperature-dependent effective mass at the VBM.
The effective mass along the $\mathrm{M}$-$\mathrm{X}$ direction, obtained from the unfolded spectral function, is shown in Fig.\,\ref{fig:spectral} (c).
The effective mass at 0\,K is 2.56\,$m_{e}$, indicated by the static-limit reference line.
The effective mass exhibits a very similar temperature dependence for both the DFT and \texttt{DeepH}-derived Hamiltonians, following power-law scalings of $T^{2.65}$ and $T^{2.69}$, respectively.
The slightly smaller effective mass predicted by \texttt{DeepH} is consistent with the correspondingly higher carrier mobility obtained from the model.
In addition, effective masses computed for larger 128-atom and 250-atom supercells at 300\,K show only minor deviations from the 54-atom result, indicating that the effective-mass prediction is robust with respect to supercell size.

\subsection{Computational Efficiency and Scalability of Deep-Learning Hamiltonian}
\label{subsec:cost}

\begin{figure}[t!]
\centering
\includegraphics[width=0.8\linewidth]{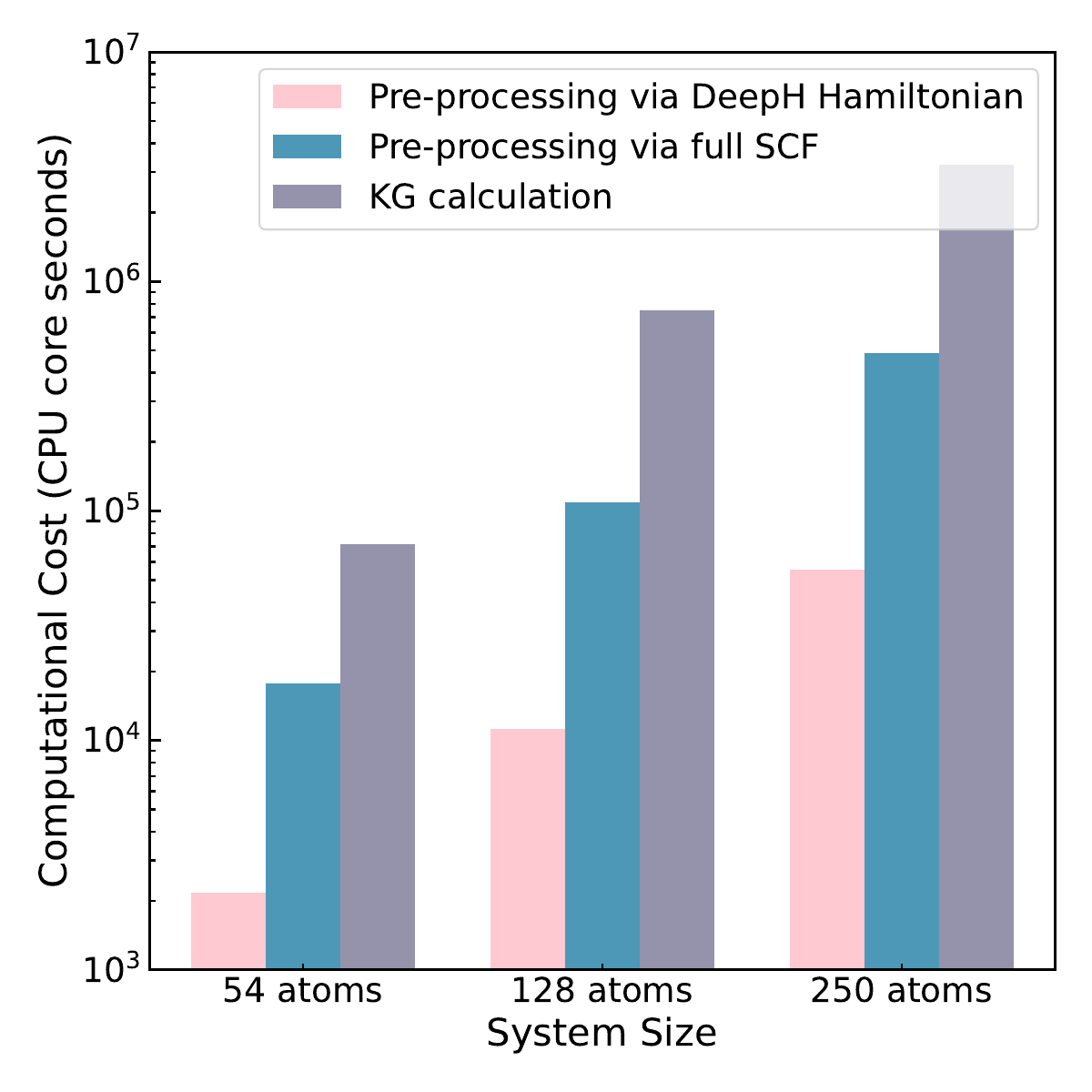}
\caption{Computational cost of KG pre-processing and KG calculation for bulk KI with different system sizes, comparing full SCF and \texttt{DeepH} Hamiltonian methods in CPU core seconds.}
\label{fig:cost}
\end{figure}

\texttt{DeepH} substantially reduces the computational cost associated with large-scale KG pre-processing.
The computational time, measured in CPU core seconds, is summarized in Fig.\,\ref{fig:cost}.
The overall workflow consists of two main steps, including pre-processing (e.g., SCF calculations) and property evaluations. Replacing conventional SCF calculations with \texttt{DeepH} Hamiltonian predictions significantly reduces the computational time required for pre-processing.
For supercells containing 54, 128, and 250 atoms, the difference in CPU time between conventional DFT and the \texttt{DeepH}-assisted approach amounts to approximately $1.6\times10^{4}$, $9.8\times10^{4}$ and $4.3\times10^{5}$ CPU core seconds, respectively.
The computational savings increase rapidly with system size, demonstrating the favorable scaling of the proposed framework for large supercells.

In addition, the pre-processing step itself is accelerated by roughly one order of magnitude.
For example, in the 128-atom supercell, the computational cost is reduced from approximately $10^{5}$ to $10^{4}$ CPU core seconds when \texttt{DeepH} is employed. For all three supercell sizes, the application of \texttt{DeepH} reduced the computational cost of the pre-processing step by approximately 90\,\%.
In contrast, subsequent property calculations, including band-structure evaluation and KG mobility spectra, remain unchanged and therefore do not directly benefit from the \texttt{DeepH} framework.
The gray region in Fig.\,\ref{fig:cost} represents the computational time required for KG calculations after the pre-processing step, which increases by roughly one order of magnitude with increasing supercell size.
Further acceleration of these post-processing steps, therefore, remains an important direction for future methodological developments.
We note that the reported timings exclude the one-time cost of training the \texttt{DeepH} model, which is approximately 30 hours on four GPUs in the present work. Once trained, however, the model can be reused for subsequent calculations of different physical properties without retraining. As a result, the training cost is amortized over multiple calculations, making the overall workflow particularly advantageous for large-scale or high-throughput applications.

\section{Computational details}
All DFT calculations in this work are performed utilizing the all-electron electronic-structure code \texttt{FHI-aims}\,\cite{blum2009ab,2015hybrid}, which employs NAOs as basis functions.
Exchange-correlation interactions are described within the generalized gradient approximation\,(GGA) using the Perdew-Burke-Ernzerhof\,(PBE) parameterization\,\cite{perdew1996}.
Although PBE underestimates the KI band gap, this error chiefly impacts the absolute band-edge energies rather than the qualitative mobility trends. The latter are largely dictated by phonon-induced scattering and should therefore remain robust, at least in the low-doping regime.
Structural relaxation, aiMD simulations, structural sampling, and anharmonicity analyses at different supercell sizes and temperatures are carried out using the \texttt{FHI-vibes}\,\cite{Knoop2020-vibes} framework, whereby Phonopy\,\cite{togo2023} is used as calculator for the harmonic force constants.
The Born-Oppenheimer aiMD trajectories are generated in the canonical (NVT) ensemble using a $2\times2\times2$ reciprocal-space grid and a time step of 4\,fs, with a total simulation time of 10\,ps (2,500\,steps).
After an initial equilibration period of approximately 2\,ps, the system reaches stable thermodynamic conditions, and the remaining trajectory frames are used to construct the training dataset for \texttt{DeepH}\,\cite{li2022deeph,gong2023,2026deeph}.
These thermally sampled structures are subsequently used for electronic band-structure calculations and to evaluate transport properties within the KG formalism.

The dataset used for training the deep-learning Hamiltonian is generated from aiMD trajectories of a 54-atom ($3\times3\times3$) KI supercell at 300\,K.
In total, 360 configurations are collected and divided into 270 training, 70 validation, and 20 test samples.
To extend the model to larger supercells and different temperatures, the pre-trained model is further fine-tuned using additional datasets.
These include 20 configurations from a 128-atom ($4\times4\times4$) KI supercell at 300\,K, as well as 50, 70, 75, and 200 configurations from 54-atom KI cells at 200, 500, 700, and 900\,K, respectively. Higher-temperature and larger-supercell points required more configurations to span the broader thermal phase space.
For each configuration, the Kohn--Sham Hamiltonian is calculated using full SCF iterations with a $2\times2\times2$ $\mathbf{k}$-point grid and the \textit{intermediate} basis set in \texttt{FHI-aims},
which corresponds to atomic orbitals \textit{s6p5d3f1} for KI. The full basis, including \textit{d} and \textit{f} functions, is included in both the reference DFT calculations and the DeepH Hamiltonian prediction. This ensures that the equivariant Hamiltonian representation handles Hamiltonian matrix elements of arbitrary orbital character up to \textit{f} symmetry.
The resulting Hamiltonian matrices are subsequently converted into the \texttt{DeepH} format for model training.
To address the ill-conditioned overlap matrix in the generalized Kohn--Sham equation arising from the non-orthogonal basis set in \texttt{FHI-aims}, the parameter \textit{basis\_threshold} is set to $10^{-4}$, such that eigenvectors of the overlap matrix below this threshold are projected out.
This procedure reduces numerical noise and improves the stability of the training process.

The neural-network model consists of three layers, with the minimum learning-rate scale set to $10^{-4}$.
The cutoff radius for the Gaussian basis functions is chosen as 13\,\AA, and the maximum momentum order $l$ in the irreducible representations is set to 5 in order to capture higher-order tensor features associated with the $f$ orbitals in KI.
All \texttt{DeepH} training and fine-tuning procedures are performed on a full node equipped with 4 NVIDIA A100 GPUs.

For electronic band structure calculations based on both DFT and \texttt{DeepH} Hamiltonians, the same $k$-point grid of $4\times4\times4$ is employed.
For the KG transport calculations, denser $\mathbf{k}$-point grids are utilized depending on the supercell size.
Specifically, $30\times30\times30$, $16\times16\times16$, and $8\times8\times8$ $\mathbf{k}$-point meshes are adopted for the 16-atom, 54/128-atom, and 250-atom ($5\times5\times5$ supercell) KI systems, respectively.
All calculations are performed using the \textit{intermediate} basis set in \texttt{FHI-aims}.
Following the methodology established in anharmonic SrTiO$_3$\cite{KG}, we adopt a carrier concentration of $n = 1\times10^{16} (cm^{-3})$ as a reference point for the electrical transport analysis, and select the flatter valence band for mobility prediction.
The frequency window $\hbar\omega$ is sampled from the band edge to 0.3 meV.
To approximate the delta function in the KG formula, a Gaussian broadening parameter $\eta$ in the range of 1-10 meV is applied.
Statistical convergence of the transport properties is achieved by averaging over 50 independent thermally sampled configurations from MD trajectories. 
Convergence of this ensemble average is verified in SM\,(Fig.\,S3). Using the existing $100$ configurations for the $54$-atom supercell at $300$~K, the mean peak mobility reaches a stable value around N = 50 and remains very close to the full 100-sample result. Specifically, at N = 50, the peak mobility is $3.09 \pm 0.24$ cm$^{2}$V$^{-1}$s$^{-1}$, while the full 100-sample value is 3.07 cm$^{2}$V$^{-1}$s$^{-1}$. This confirms that 50 samples are sufficient and statistically reliable for the reported peak mobility.

The temperature-dependent spectral functions and band-gap renormalization are obtained by unfolding the electronic band structures of supercell aiMD snapshots into the primitive-cell Brillouin zone, followed by an ensemble average over 50 independent configurations using a $4\times4\times4$ $\mathbf{k}$-point grid.
The \texttt{DeepH}-predicted Hamiltonian is used in the same procedure to evaluate the spectral functions without performing additional SCF iterations.
To extract the CBM or VBM peak of the spectral function, a Lorentzian\,\cite{giustino2017} line shape is fitted at the corresponding $\mathbf{k}$ point.
However, special care is required at elevated temperatures due to the significant spectral broadening.
Near the VBM peak around the $\mathrm{M}$ point, the effective mass is obtained by performing a parabolic fit using four $\mathbf{k}$ points along the high-symmetry $\mathrm{M}-\Gamma$ direction.

\vspace{2em}
\section{Conclusion and outlook}
In this work, we establish an AI-assisted framework for non-perturbative transport calculations in strongly anharmonic materials by combining the aiKG formalism with the deep-learning Hamiltonian model \texttt{DeepH}.
Using bulk KI as a representative benchmark, we show that the Kohn--Sham Hamiltonian can be predicted with high fidelity even in the strongly anharmonic regime, enabling reliable large-supercell electronic-structure calculations without repeated self-consistent first-principles evaluations.
By explicitly incorporating fully anharmonic lattice dynamics, the proposed approach reproduces temperature-dependent carrier mobility, spectral functions, and effective masses in close agreement with DFT.
These results demonstrate that deep-learning Hamiltonians provide a methodology for KI-class strongly anharmonic systems and a template for future extensions to other material classes once corresponding training sets are available.

At the same time, the present results highlight an important challenge for transport-oriented Hamiltonian learning. Even very small errors in the predicted Hamiltonian can be amplified during the solution of the generalized Kohn--Sham equation, leading to noticeably larger deviations in eigenvalues and, consequently, in transport observables that are highly sensitive to the underlying electronic structure. This issue becomes particularly relevant at finite temperatures, where ensemble averaging over thermally sampled configurations can further magnify configuration-dependent discrepancies. 
This amplification should become more acute in larger or more complex systems. 
Several mitigation strategies follow from the cascade analysis above.
These include (i) augmenting the training loss with eigenvalue or band-velocity penalties that target band-edge states, so that the quantities most relevant to transport are weighted preferentially during training; (ii) training multi-model ensembles to provide per-configuration uncertainty estimates and to flag high-uncertainty configurations for full-DFT recomputation; (iii) performing systematic
per-supercell validation against at least one DFT reference configuration whenever the eigenvalue MAE approaches $\sim\!10$~meV; and (iv) developing transport-aware loss functions that directly penalize
errors in the computed conductivity spectrum. 

In addition, the computational cost of generating training data could be substantially reduced by replacing aiMD with molecular dynamics driven by MLIP-MD for configurational sampling. Combined with active-learning strategies, only high-uncertainty configurations need to be recomputed at the DFT level, enabling efficient exploration of larger supercells, longer trajectories, and more complex free-energy landscapes while maintaining first-principles accuracy. Together, these transport-oriented learning strategies provide a scalable route toward first-principles transport studies of strongly anharmonic systems and, ultimately, high-throughput screening of functional materials with complex lattice dynamics. The present work on strongly anharmonic KI therefore not only provides a practical tool for accelerating non-perturbative transport calculations but also establishes a foundation for the development of next-generation transport-oriented Hamiltonian learning frameworks.

\section{Acknowledgments}
This work was supported by the ERC Advanced Grant TEC1p (the European Research Council (ERC) Horizon2020 research and innovation programme, grant agreement No. 740233). Boheng Zhao, Yang Li, and Yong Xu were supported by the Basic Science Center Project of NSFC (grant no. 52388201), the National Key Basic Research and Development Program of China (grants no. 2024YFA1409100 and no. 2023YFA1406400), the National Natural Science Foundation of China (grants no. 12334003, no. 12421004, and no. 12361141826), and Fundamental and Interdisciplinary Disciplines Breakthrough Plan of the Ministry of Education of China (JYB2025XDXM408). Kisung Kang was supported by the Korea Institute of Science and Technology Information (KISTI, grant no. KSC-2024-CRE-0399). We thank Jingkai Quan, Minye Zhang, and Shuo Zhao (NOMAD group) for discussions, and Zechen Tang and Zixu Wang for assistance with the DeepH method.
\FloatBarrier

\bibliography{sample}

@article{ASI,
  title={Atomic Simulation Interface (ASI): application programming interface for electronic structure codes},
  author={Stishenko, Pavel V and Keal, Thomas W and Woodley, Scott M and Blum, Volker and Hourahine, Benjamin and Maurer, Reinhard J and Logsdail, Andrew J},
  journal={Journal of Open Source Software},
  volume={8},
  number={85},
  year={2023}
}

@article{KG,
  title = {Carrier mobility of strongly anharmonic materials from first principles},
  author = {Quan, Jingkai and Carbogno, Christian and Scheffler, Matthias},
  journal = {Phys. Rev. B},
  volume = {110},
  issue = {23},
  pages = {235202},
  numpages = {19},
  year = {2024},
  month = {Dec},
  publisher = {American Physical Society},
}

@article{ioffe,
  title={Non-crystalline, amorphous, and liquid electronic semiconductors},
  author={Ioffe, AF and Regel, AR},
  booktitle={Progress in semiconductors},
  pages={237--291},
  year={1960}
}

@article{green,
  title={Markoff random processes and the statistical mechanics of time-dependent phenomena. II. Irreversible processes in fluids},
  author={Green, Melville S},
  journal={The Journal of chemical physics},
  volume={22},
  number={3},
  pages={398--413},
  year={1954},
  publisher={American Institute of Physics}
}

@article{2026deeph,
  title={DeepH-pack: a general-purpose neural network package for deep-learning electronic structure calculations},
  author={Li, Yang and Wang, Yanzhen and Zhao, Boheng and Gong, Xiaoxun and Wang, Yuxiang and Tang, Zechen and Wang, Zixu and Yuan, Zilong and Li, Jialin and Sun, Minghui and others},
  journal={npj Computational Materials},
  year={2026},
  publisher={Nature Publishing Group UK London}
}

@article{hegde2017,
  title={Machine-learned approximations to density functional theory hamiltonians},
  author={Hegde, Ganesh and Bowen, R Chris},
  journal={Scientific reports},
  volume={7},
  number={1},
  pages={42669},
  year={2017},
  publisher={Nature Publishing Group UK London}
}

@article{schutt2019,
  title={Unifying machine learning and quantum chemistry with a deep neural network for molecular wavefunctions},
  author={Sch{\"u}tt, Kristof T and Gastegger, Michael and Tkatchenko, Alexandre and M{\"u}ller, K-R and Maurer, Reinhard J},
  journal={Nature communications},
  volume={10},
  number={1},
  pages={5024},
  year={2019},
  publisher={Nature Publishing Group UK London}
}

@article{purcell,
  title={Accelerating materials-space exploration for thermal insulators by mapping materials properties via artificial intelligence},
  author={Purcell, Thomas AR and Scheffler, Matthias and Ghiringhelli, Luca M and Carbogno, Christian},
  journal={npj computational materials},
  volume={9},
  number={1},
  pages={112},
  year={2023},
  publisher={Nature Publishing Group UK London}
}

@article{giustino,
  title={Electron-phonon interactions from first principles},
  author={Giustino, Feliciano},
  journal={Reviews of Modern Physics},
  volume={89},
  number={1},
  pages={015003},
  year={2017},
  publisher={APS}
}

@article{PRB224310,
  title = {Many-body perturbation theory approach to the electron-phonon interaction with density-functional theory as a starting point},
  author = {Marini, Andrea and Ponc\'e, S. and Gonze, X.},
  journal = {Phys. Rev. B},
  volume = {91},
  issue = {22},
  pages = {224310},
  numpages = {22},
  year = {2015},
  month = {Jun},
  publisher = {American Physical Society}
}

@article{babadi2017,
  title={Theory of parametrically amplified electron-phonon superconductivity},
  author={Babadi, Mehrtash and Knap, Michael and Martin, Ivar and Refael, Gil and Demler, Eugene},
  journal={Physical Review B},
  volume={96},
  number={1},
  pages={014512},
  year={2017},
  publisher={APS}
}

@article{chang2018,
  title={Anharmoncity and low thermal conductivity in thermoelectrics},
  author={Chang, Cheng and Zhao, Li-Dong},
  journal={Materials Today Physics},
  volume={4},
  pages={50--57},
  year={2018},
  publisher={Elsevier}
}

@article{wu2020,
  title={Strong lattice anharmonicity securing intrinsically low lattice thermal conductivity and high performance thermoelectric SnSb2Te4 via Se alloying},
  author={Wu, Hong and Lu, Xu and Wang, Guoyu and Peng, Kunling and Zhang, Bin and Chen, Yongjin and Gong, Xiangnan and Tang, Xiaodan and Zhang, Xuemei and Feng, Zhenzhen and others},
  journal={Nano Energy},
  volume={76},
  pages={105084},
  year={2020},
  publisher={Elsevier}
}

@article{song2023,
  title={Strong anharmonic phonon scattering and superior thermoelectric properties of Li2NaBi},
  author={Song, Xuhao and Zhao, Yinchang and Ni, Jun and Meng, Sheng and Dai, Zhenhong},
  journal={Materials Today Physics},
  volume={31},
  pages={100990},
  year={2023},
  publisher={Elsevier}
}

@article{simoncell,
  title={Unified theory of thermal transport in crystals and glasses},
  author={Simoncelli, Michele and Marzari, Nicola and Mauri, Francesco},
  journal={Nature Physics},
  volume={15},
  number={8},
  pages={809--813},
  year={2019},
  publisher={Nature Publishing Group UK London}
}

@book{peierls,
  title={Zur kinetischen theorie der w{\"a}rmeleitung in kristallen},
  author={Peierls, Rudolf Ernst},
  year={1929},
  publisher={JA Barth}
}

@article{gong2023,
  title={General framework for E (3)-equivariant neural network representation of density functional theory Hamiltonian},
  author={Gong, Xiaoxun and Li, He and Zou, Nianlong and Xu, Runzhang and Duan, Wenhui and Xu, Yong},
  journal={Nature Communications},
  volume={14},
  number={1},
  pages={2848},
  year={2023},
  publisher={Nature Publishing Group UK London}
}

@article{KI1967,
  title={The temperature dependence of atomic displacements in potassium iodide},
  author={Pearman, GT and Tompson, CW},
  journal={Journal of Physics and Chemistry of Solids},
  volume={28},
  number={2},
  pages={261--266},
  year={1967},
  publisher={Elsevier}
}

@article{blum2009ab,
  title={Ab initio molecular simulations with numeric atom-centered orbitals},
  author={Blum, Volker and Gehrke, Ralf and Hanke, Felix and Havu, Paula and Havu, Ville and Ren, Xinguo and Reuter, Karsten and Scheffler, Matthias},
  journal={Computer Physics Communications},
  volume={180},
  number={11},
  pages={2175--2196},
  year={2009},
  publisher={Elsevier}
}

@article{2015hybrid,
  title={Hybrid functionals for large periodic systems in an all-electron, numeric atom-centered basis framework},
  author={Levchenko, Sergey V and Ren, Xinguo and Wieferink, J{\"u}rgen and Johanni, Rainer and Rinke, Patrick and Blum, Volker and Scheffler, Matthias},
  journal={Computer Physics Communications},
  volume={192},
  pages={60--69},
  year={2015},
  publisher={Elsevier}
}

@article{li2022deeph,
  title={Deep-learning density functional theory Hamiltonian for efficient ab initio electronic-structure calculation},
  author={Li, He and Wang, Zun and Zou, Nianlong and Ye, Meng and Xu, Runzhang and Gong, Xiaoxun and Duan, Wenhui and Xu, Yong},
  journal={Nature Computational Science},
  volume={2},
  number={6},
  pages={367--377},
  year={2022},
  publisher={Nature Publishing Group US New York}
}

@article{2023atomic,
  title={Atomic Simulation Interface (ASI): application programming interface for electronic structure codes},
  author={Stishenko, Pavel V and Keal, Thomas W and Woodley, Scott M and Blum, Volker and Hourahine, Benjamin and Maurer, Reinhard J and Logsdail, Andrew J},
  journal={Journal of Open Source Software},
  volume={8},
  number={85},
  year={2023}
}

@article{holst2011,
  title={Electronic transport coefficients from ab initio simulations and application to dense liquid hydrogen},
  author={Holst, Bastian and French, Martin and Redmer, Ronald},
  journal={Physical Review B—Condensed Matter and Materials Physics},
  volume={83},
  number={23},
  pages={235120},
  year={2011},
  publisher={APS}
}

@article{kubo1957,
  title={Statistical-mechanical theory of irreversible processes. I. General theory and simple applications to magnetic and conduction problems},
  author={Kubo, Ryogo},
  journal={Journal of the physical society of Japan},
  volume={12},
  number={6},
  pages={570--586},
  year={1957},
  publisher={The Physical Society of Japan}
}

@article{2022optical,
  title={Optical conductivity of an anharmonic large polaron gas at weak coupling},
  author={Houtput, Matthew and Tempere, Jacques},
  journal={Physical Review B},
  volume={106},
  number={21},
  pages={214315},
  year={2022},
  publisher={APS}
}

@article{gigli2024,
  title={Mechanism of charge transport in lithium thiophosphate},
  author={Gigli, Lorenzo and Tisi, Davide and Grasselli, Federico and Ceriotti, Michele},
  journal={Chemistry of Materials},
  volume={36},
  number={3},
  pages={1482--1496},
  year={2024},
  publisher={ACS Publications}
}

@article{greenwood1958,
  title={The Boltzmann equation in the theory of electrical conduction in metals},
  author={Greenwood, DA},
  journal={Proceedings of the Physical Society},
  volume={71},
  number={4},
  pages={585},
  year={1958},
  publisher={IOP Publishing}
}

@article{knuth2015,
  title={All-electron formalism for total energy strain derivatives and stress tensor components for numeric atom-centered orbitals},
  author={Knuth, Franz and Carbogno, Christian and Atalla, Viktor and Blum, Volker and Scheffler, Matthias},
  journal={Computer Physics Communications},
  volume={190},
  pages={33--50},
  year={2015},
  publisher={Elsevier}
}

@article{shang2017,
  title={Lattice dynamics calculations based on density-functional perturbation theory in real space},
  author={Shang, Honghui and Carbogno, Christian and Rinke, Patrick and Scheffler, Matthias},
  journal={Computer Physics Communications},
  volume={215},
  pages={26--46},
  year={2017},
  publisher={Elsevier}
}

@book{bruus2004,
  title={Many-body quantum theory in condensed matter physics: an introduction},
  author={Bruus, Henrik and Flensberg, Karsten},
  year={2004},
  publisher={Oxford university press}
}

@book{coleman2015,
  title={Introduction to many-body physics},
  author={Coleman, Piers},
  year={2015},
  publisher={Cambridge University Press}
}

@article{zacharias2020,
  title={Fully anharmonic nonperturbative theory of vibronically renormalized electronic band structures},
  author={Zacharias, Marios and Scheffler, Matthias and Carbogno, Christian},
  journal={Physical Review B},
  volume={102},
  number={4},
  pages={045126},
  year={2020},
  publisher={APS}
}

@article{quan2025,
  title = {Efficient band structure unfolding with atom-centered orbitals: General theory and application},
  author = {Quan, Jingkai and Rybin, Nikita and Scheffler, Matthias and Carbogno, Christian},
  journal = {Phys. Rev. B},
  volume = {113},
  issue = {8},
  pages = {085112},
  numpages = {14},
  year = {2026},
  month = {Feb},
  publisher = {American Physical Society}
}

@article{giustino2017,
  title={Electron-phonon interactions from first principles},
  author={Giustino, Feliciano},
  journal={Reviews of Modern Physics},
  volume={89},
  number={1},
  pages={015003},
  year={2017},
  publisher={APS}
}

@article{carbogno2017,
  title={Ab initio Green-Kubo approach for the thermal conductivity of solids},
  author={Carbogno, Christian and Ramprasad, Rampi and Scheffler, Matthias},
  journal={Physical review letters},
  volume={118},
  number={17},
  pages={175901},
  year={2017},
  publisher={APS}
}

@article{knoop2023,
  title={Ab initio Green-Kubo simulations of heat transport in solids: Method and implementation},
  author={Knoop, Florian and Scheffler, Matthias and Carbogno, Christian},
  journal={Physical Review B},
  volume={107},
  number={22},
  pages={224304},
  year={2023},
  publisher={APS}
}

@article{snyder2008,
  title={Complex thermoelectric materials},
  author={Snyder, G Jeffrey and Toberer, Eric S},
  journal={Nature materials},
  volume={7},
  number={2},
  pages={105--114},
  year={2008},
  publisher={Nature Publishing Group UK London}
}

@article{he2017,
  title={Advances in thermoelectric materials research: Looking back and moving forward},
  author={He, Jian and Tritt, Terry M},
  journal={Science},
  volume={357},
  number={6358},
  pages={eaak9997},
  year={2017},
  publisher={American Association for the Advancement of Science}
}

@article{zevalkink2018,
  title={A practical field guide to thermoelectrics: Fundamentals, synthesis, and characterization},
  author={Zevalkink, Alex and Smiadak, David M and Blackburn, Jeff L and Ferguson, Andrew J and Chabinyc, Michael L and Delaire, Olivier and Wang, Jian and Kovnir, Kirill and Martin, Joshua and Schelhas, Laura T and others},
  journal={Applied Physics Reviews},
  volume={5},
  number={2},
  year={2018},
  publisher={AIP Publishing}
}

@article{xia2020,
  title={Particlelike phonon propagation dominates ultralow lattice thermal conductivity in crystalline Tl 3 VSe 4},
  author={Xia, Yi and Pal, Koushik and He, Jiangang and Ozoli{\c{n}}{\v{s}}, Vidvuds and Wolverton, Chris},
  journal={Physical Review Letters},
  volume={124},
  number={6},
  pages={065901},
  year={2020},
  publisher={APS}
}

@article{knoop2020,
  title={Anharmonicity measure for materials},
  author={Knoop, Florian and Purcell, Thomas AR and Scheffler, Matthias and Carbogno, Christian},
  journal={Physical Review Materials},
  volume={4},
  number={8},
  pages={083809},
  year={2020},
  publisher={APS}
}

@article{baroni2001,
  title={Phonons and related crystal properties from density-functional perturbation theory},
  author={Baroni, Stefano and De Gironcoli, Stefano and Dal Corso, Andrea and Giannozzi, Paolo},
  journal={Reviews of modern Physics},
  volume={73},
  number={2},
  pages={515},
  year={2001},
  publisher={APS}
}

@article{ponce2021,
  title={First-principles predictions of Hall and drift mobilities in semiconductors},
  author={Ponc{\'e}, Samuel and Macheda, Francesco and Margine, Elena Roxana and Marzari, Nicola and Bonini, Nicola and Giustino, Feliciano},
  journal={Physical Review Research},
  volume={3},
  number={4},
  pages={043022},
  year={2021},
  publisher={APS}
}

@article{errea2014,
  title={Anharmonic free energies and phonon dispersions from the stochastic self-consistent harmonic approximation: Application to platinum and palladium hydrides},
  author={Errea, Ion and Calandra, Matteo and Mauri, Francesco},
  journal={Physical Review B},
  volume={89},
  number={6},
  pages={064302},
  year={2014},
  publisher={APS}
}

@article{shakouri2011,
  title={Recent developments in semiconductor thermoelectric physics and materials},
  author={Shakouri, Ali},
  journal={Annual review of materials research},
  volume={41},
  number={1},
  pages={399--431},
  year={2011},
  publisher={Annual Reviews}
}

@article{li2012,
  title={Thermal conductivity of bulk and nanowire Mg 2 Si x Sn 1- x alloys from first principles},
  author={Li, Wu and Lindsay, Lucas and Broido, David A and Stewart, Derek A and Mingo, Natalio},
  journal={Physical Review B—Condensed Matter and Materials Physics},
  volume={86},
  number={17},
  pages={174307},
  year={2012},
  publisher={APS}
}

@article{lanzara2001,
  title={Evidence for ubiquitous strong electron--phonon coupling in high-temperature superconductors},
  author={Lanzara, A and Bogdanov, PV and Zhou, XJ and Kellar, SA and Feng, DL and Lu, ED and Yoshida, Teppei and Eisaki, H and Fujimori, Atsushi and Kishio, Kohji and others},
  journal={Nature},
  volume={412},
  number={6846},
  pages={510--514},
  year={2001},
  publisher={Nature Publishing Group UK London}
}

@article{wright2016,
  title={Electron--phonon coupling in hybrid lead halide perovskites},
  author={Wright, Adam D and Verdi, Carla and Milot, Rebecca L and Eperon, Giles E and P{\'e}rez-Osorio, Miguel A and Snaith, Henry J and Giustino, Feliciano and Johnston, Michael B and Herz, Laura M},
  journal={Nature communications},
  volume={7},
  number={1},
  pages={11755},
  year={2016},
  publisher={Nature Publishing Group UK London}
}

@article{zhang2025,
  title={Enhanced long-range quadrupole effects in 2D MSi2N4: impacts on electric and thermal transport},
  author={Zhang, Juan and Gong, Jiayi and Chen, Hongyu and Peng, Lei and Shao, Hezhu and Cen, Yan and Zhuang, Jun and Zhu, Heyuan and Zhou, Jinjian and Zhang, Hao},
  journal={npj Computational Materials},
  volume={11},
  number={1},
  pages={166},
  year={2025},
  publisher={Nature Publishing Group UK London}
}

@article{beltukov2013,
  title={Ioffe-Regel criterion and diffusion of vibrations in random lattices},
  author={Beltukov, YM and Kozub, VI and Parshin, DA},
  journal={Physical Review B—Condensed Matter and Materials Physics},
  volume={87},
  number={13},
  pages={134203},
  year={2013},
  publisher={APS}
}

@article{tadano2018,
  title={Quartic anharmonicity of rattlers and its effect on lattice thermal conductivity of clathrates from first principles},
  author={Tadano, Terumasa and Tsuneyuki, Shinji},
  journal={Physical review letters},
  volume={120},
  number={10},
  pages={105901},
  year={2018},
  publisher={APS}
}

@article{feng2017,
  title={Four-phonon scattering significantly reduces intrinsic thermal conductivity of solids},
  author={Feng, Tianli and Lindsay, Lucas and Ruan, Xiulin},
  journal={Physical Review B},
  volume={96},
  number={16},
  pages={161201},
  year={2017},
  publisher={APS}
}

@article{wang2021,
  title={Revisiting phonon transport in perovskite SrTiO 3: Anharmonic phonon renormalization and four-phonon scattering},
  author={Wang, Qi and Zeng, Zezhu and Chen, Yue},
  journal={Physical Review B},
  volume={104},
  number={23},
  pages={235205},
  year={2021},
  publisher={APS}
}

@article{ouyang2022,
  title={Accurate description of high-order phonon anharmonicity and lattice thermal conductivity from molecular dynamics simulations with machine learning potential},
  author={Ouyang, Yulou and Yu, Cuiqian and He, Jia and Jiang, Pengfei and Ren, Weijun and Chen, Jie},
  journal={Physical Review B},
  volume={105},
  number={11},
  pages={115202},
  year={2022},
  publisher={APS}
}

@article{qian2025,
  title={Equivariant electronic Hamiltonian prediction with many-body message passing},
  author={Qian, Chen and Vitartas, Valdas and Kermode, James R and Maurer, Reinhard J},
  journal={npj Computational Materials},
  year={2026},
  publisher={Nature Publishing Group UK London}
}

@article{dolling1966,
  title={Normal vibrations of potassium iodide},
  author={Dolling, G and Cowley, RA and Schittenhelm, C and Thorson, IM},
  journal={Physical Review},
  volume={147},
  number={2},
  pages={577},
  year={1966},
  publisher={APS}
}

@article{freville2023,
  title={Equation of state of KI up to 150 GPa},
  author={Fr{\'e}ville, Robin and Dewaele, Agn{\`e}s and Guignot, Nicolas and Garbarino, Gaston and Mezouar, Mohamed},
  journal={High Pressure Research},
  volume={43},
  number={3},
  pages={251--261},
  year={2023},
  publisher={Taylor \& Francis}
}

@article{zhou2018,
  title = {Electron-Phonon Scattering in the Presence of Soft Modes and Electron Mobility in ${\mathrm{SrTiO}}_{3}$ Perovskite from First Principles},
  author = {Zhou, Jin-Jian and Hellman, Olle and Bernardi, Marco},
  journal = {Phys. Rev. Lett.},
  volume = {121},
  issue = {22},
  pages = {226603},
  numpages = {6},
  year = {2018},
  month = {Nov},
  publisher = {American Physical Society},
}

@article{ponce2018,
  title = {Towards predictive many-body calculations of phonon-limited carrier mobilities in semiconductors},
  author = {Ponc\'e, Samuel and Margine, Elena R. and Giustino, Feliciano},
  journal = {Phys. Rev. B},
  volume = {97},
  issue = {12},
  pages = {121201},
  numpages = {5},
  year = {2018},
  month = {Mar},
  publisher = {American Physical Society},
}

@book{drude1900,
  title={Annalen der Physik},
  author={Drude, Paul and Wien, Wilhelm and Gr{\"u}neisen, Eduard August and Planck, Max},
  volume={3},
  year={1900},
  publisher={JA Barth}
}

@article{willis2013,
  title={A generalized Drude model for doped silicon at terahertz frequencies derived from microscopic transport simulation},
  author={Willis, KJ and Hagness, SC and Knezevic, I},
  journal={Applied Physics Letters},
  volume={102},
  number={12},
  year={2013},
  publisher={AIP Publishing}
}

@article{ponce2020,
  title={First-principles calculations of charge carrier mobility and conductivity in bulk semiconductors and two-dimensional materials},
  author={Ponc{\'e}, Samuel and Li, Wenbin and Reichardt, Sven and Giustino, Feliciano},
  journal={Reports on Progress in Physics},
  volume={83},
  number={3},
  pages={036501},
  year={2020},
  publisher={IOP Publishing}
}

@article{li2015,
  title={Electrical transport limited by electron-phonon coupling from Boltzmann transport equation: An ab initio study of Si, Al, and MoS 2},
  author={Li, Wu},
  journal={Physical Review B},
  volume={92},
  number={7},
  pages={075405},
  year={2015},
  publisher={APS}
}

@article{kang2025,
  title={Accelerating the training and improving the reliability of machine-learned interatomic potentials for strongly anharmonic materials through active learning},
  author={Kang, Kisung and Purcell, Thomas AR and Carbogno, Christian and Scheffler, Matthias},
  journal={Physical Review Materials},
  volume={9},
  number={6},
  pages={063801},
  year={2025},
  publisher={APS}
}

@article{prodan2005,
  title={Nearsightedness of electronic matter},
  author={Prodan, Emil and Kohn, Walter},
  journal={Proceedings of the National Academy of Sciences},
  volume={102},
  number={33},
  pages={11635--11638},
  year={2005},
  publisher={National Academy of Sciences}
}

@inproceedings{passaro2023,
  title={Reducing SO(3) convolutions to SO(2) for efficient equivariant GNNs},
  author={Passaro, Saro and Zitnick, C Lawrence},
  booktitle={International conference on machine learning},
  pages={27420--27438},
  year={2023},
  organization={PMLR}
}

@article{2019machine,
  title={Machine learning interatomic potentials as emerging tools for materials science},
  author={Deringer, Volker L and Caro, Miguel A and Cs{\'a}nyi, G{\'a}bor},
  journal={Advanced Materials},
  volume={31},
  number={46},
  pages={1902765},
  year={2019},
  publisher={Wiley Online Library}
}

@article{lorenz2004,
  title={Representing high-dimensional potential-energy surfaces for reactions at surfaces by neural networks},
  author={Lorenz, S{\"o}nke and Gro{\ss}, Axel and Scheffler, Matthias},
  journal={Chemical Physics Letters},
  volume={395},
  number={4-6},
  pages={210--215},
  year={2004},
  publisher={Elsevier}
}

@article{perdew1996,
  title={Generalized gradient approximation made simple},
  author={Perdew, John P and Burke, Kieron and Ernzerhof, Matthias},
  journal={Physical review letters},
  volume={77},
  number={18},
  pages={3865},
  year={1996},
  publisher={APS}
}

@article{Knoop2020-vibes,
  doi = {10.21105/joss.02671},
  url = {https://doi.org/10.21105/joss.02671},
  year = {2020},
  publisher = {The Open Journal},
  volume = {5},
  number = {56},
  pages = {2671},
  author = {Florian Knoop and Thomas A. R. Purcell and Matthias Scheffler and Christian Carbogno},
  title = {FHI-vibes: _Ab Initio_ Vibrational Simulations},
  journal = {Journal of Open Source Software}
}

@article{togo2023,
  title={Implementation strategies in phonopy and phono3py},
  author={Togo, Atsushi and Chaput, Laurent and Tadano, Terumasa and Tanaka, Isao},
  journal={Journal of Physics: Condensed Matter},
  volume={35},
  number={35},
  pages={353001},
  year={2023},
  publisher={IOP Publishing}
}

@article{bardeen1950,
  author  = {Bardeen, J. and Shockley, W.},
  title   = {Deformation Potentials and Mobilities in Non-Polar Crystals},
  journal = {Phys. Rev.},
  volume  = {80},
  pages   = {72},
  year    = {1950},
  doi     = {10.1103/PhysRev.80.72}
}

@article{frohlich1954,
  author  = {Fr\"ohlich, H.},
  title   = {Electrons in Lattice Fields},
  journal = {Adv. Phys.},
  volume  = {3},
  pages   = {325},
  year    = {1954},
  doi     = {10.1080/00018735400101213}
}

@article{renormalization,
  title={A PRIMARY RESEARCH OF THE PHONON RENORMALIZATION OF NUCLEAR FIELD THEORY},
  author={Ben-hao, Sa and Xi-zhen, Zhang and Zhu-xia, Li and Yi-jin, Shi},
  journal={Chinese Physics C},
  volume={4},
  number={3},
  pages={398--400},
  year={1980},
  publisher={Chinese Physics C}
}

@article{KI-experiment1,
  title = {Optical Absorption Spectra of the Alkali Halides at 10\ifmmode^\circ\else\textdegree\fi{}K},
  author = {Teegarden, K. and Baldini, G.},
  journal = {Phys. Rev.},
  volume = {155},
  issue = {3},
  pages = {896--907},
  numpages = {0},
  year = {1967},
  month = {Mar},
  publisher = {American Physical Society},
}

@article{KI-experiment2,
  title = {Two-Quantum Absorption Spectrum of KI and CsI},
  author = {Hopfield, J. J. and Worlock, J. M.},
  journal = {Phys. Rev.},
  volume = {137},
  issue = {5A},
  pages = {A1455--A1464},
  numpages = {0},
  year = {1965},
  month = {Mar},
  publisher = {American Physical Society},
}

@article{Redmer,
  title={Electronic transport coefficients from density functional theory across the plasma plane},
  author={French, Martin and R{\"o}pke, Gerd and Sch{\"o}rner, Maximilian and Bethkenhagen, Mandy and Desjarlais, Michael P and Redmer, Ronald},
  journal={Physical Review E},
  volume={105},
  number={6},
  pages={065204},
  year={2022},
  publisher={APS}
}

\end{document}